\documentclass[aps,prd,showpacs,notitlepage,nofootinbib,superscriptaddress,floatfix,showkeys,twocolumn]{revtex4-1}

\usepackage{blindtext}
\usepackage{hyperref}
\usepackage{amsmath,amssymb}
\usepackage{float}
\usepackage{microtype}
\usepackage{ragged2e}  

\usepackage{graphicx}
\usepackage{bm}
\usepackage{latexsym}
\usepackage{epsfig}
\usepackage{psfrag}
\usepackage[dvipsnames]{xcolor}
\usepackage{subcaption}
\usepackage{graphicx}
\usepackage{multirow}
\usepackage{amssymb}
\usepackage{amsmath}
\usepackage{bm}
\usepackage{latexsym}
\usepackage{epsfig}
\usepackage{psfrag}
\usepackage[normalem]{ulem}
\usepackage{textcomp}
\usepackage{color}
\usepackage{pstricks}
\usepackage[utf8]{inputenc}
\usepackage{comment}
\usepackage{hyperref}
\usepackage[mathlines]{lineno}
\usepackage{hyperref}
\usepackage[all]{hypcap}
\hypersetup{  
    colorlinks=true,
    linkcolor=blue,
    filecolor=red,      
    urlcolor=blue,
    citecolor=blue,
    }

\def\HI{$\textrm{H}\textsc{i}$}

\begin{document}
\title{
Redshift-space 21-cm bispectrum multipoles as an SKA-era gravity test in the post-reionization Universe
}

\author{Sourav Pal}
\email{soupal1729@gmail.com }
\affiliation{\,Physics and Applied Mathematics Unit, Indian Statistical Institute, 203 B.T. Road, Kolkata 700108, India}

\author{Debanjan Sarkar}
\email{debanjan.sarkar@mcgill.ca}
\affiliation{\,Department of Physics and Trottier Space Institute, McGill University, QC H3A 2T8, Canada}
\affiliation{\,Ciela—Montreal Institute for Astrophysical Data Analysis and Machine Learning, QC H2V0B3, Canada}


\begin{abstract}

The redshifted 21-cm line from neutral hydrogen ($\textrm{H}\textsc{i}$) enables volumetric intensity mapping of large-scale structure in the post-reionization Universe. In anticipation of \texttt{SKA-MID}'s wide redshift coverage and high signal-to-noise clustering measurements, we study the redshift-space 21-cm bispectrum and its spherical-harmonic multipoles as probes of anisotropic non-linear structure formation and departures from General Relativity. Using a tree-level perturbative description for the 21-cm brightness-temperature field in redshift space, and adopting the Hu--Sawicki $f(R)$ model as a representative modified-gravity scenario, we forecast the detectability of configuration-dependent signatures with an \texttt{SKA-MID}--like survey. We derive the bispectrum-multipole covariance including sample variance and thermal noise and evaluate the expected signal-to-noise of deviations relative to $\Lambda$CDM. We find that the observable information is dominated by the lowest multipoles, while higher-order modes are strongly suppressed. This concentration in the lowest multipoles is well matched to \texttt{SKA-MID} sensitivity and to the quasi-linear modes that are expected to remain accessible in practice. The strongest modified-gravity sensitivity arises from squeezed and stretched triangle configurations on quasi-linear scales, where scale-dependent growth enhances the bispectrum relative to the total variance. Our results position 21-cm bispectrum multipoles as a practical, SKA-ready observable for testing gravity beyond $\Lambda$CDM in the post-reionization epoch.

\end{abstract}

\maketitle

\section{Introduction}\label{sec:introduction}
\label{sec:introduction}

The redshifted 21-cm hyperfine transition of neutral hydrogen (\HI) provides a direct, three-dimensional tracer of large-scale structure over an exceptionally wide span of cosmic time~\citep{Bharadwaj:2000av,Pritchard:2008da,pritchard201221,Furlanetto:2009qk,Liu:2012xy,Munoz:2020itp,Furlanetto:2006jb}. In the post-reionization era ($z \lesssim 6$), most of the intergalactic medium is ionized \citep{SDSS:2001tew,Fan:2006dp,Fan:2005es} and the remaining \HI\ is hosted predominantly in dense, self-shielded systems (e.g.\ galaxies and damped Lyman-$\alpha$ absorbers)~\citep{ Storrie-Lombardi:2000uit, Santos:2007dn,Villaescusa-Navarro:2014cma,Nunhokee:2025jbn,CHIME:2025cee}. 21-cm intensity mapping (IM) targets the aggregate, unresolved \HI\ emission, yielding a continuous brightness-temperature field that efficiently maps extremely large comoving volumes without resolving individual galaxies~\citep{battye2012bingo,Bull:2014rha,Camera:2015yqa,Liu:2019jiy}. This ``large-volume, low-resolution'' advantage is particularly compelling for precision cosmology: it enables high-statistics measurements of clustering over wide redshift ranges that are difficult to access with traditional galaxy surveys \citep{Bharadwaj:2000av, Wyithe:2008mv, Wyithe:2007rq}.

The IM programme has now entered a stage where end-to-end analyses on real data are becoming increasingly mature. Recent measurements and limits have begun to constrain the post-EoR \HI\ power spectrum~\citep{Paul:2023yrr,CHIME:2025cee} and its cross-correlations~\citep{Masui:2012zc, Anderson:2017ert, eBOSS:2021ebm, Cunnington:2022uzo, CHIME:2023til}, demonstrating the viability of IM pipelines and sharpening the requirements on foreground mitigation, calibration, and systematics control~\citep{Liu:2019awk,CHIME:2022kvg,MeerKLASS:2024ypg}. Looking ahead, \texttt{SKA-MID} is poised to be a transformative facility for post-EoR 21-cm cosmology~\citep{cosmicVisions21cm:2018rfq,SKA:2018ckk,Weltman:2018zrl}. Its large collecting area, wide sky coverage, and sensitivity over the intermediate-redshift window ($z \sim 0.3$--$3$) make it uniquely suited to push beyond first detections toward precision measurements of linear and non-linear clustering statistics.

The cosmological information in 21-cm maps has so far been proposed to extract primarily through two-point statistics, or equivalently the power spectrum~\citep{Wyithe:2008mv,Sarkar:2016lvb,Padmanabhan:2016fgy,Obuljen:2017jiy,Sarkar:2018gcb,Sarkar:2019nak,Berti:2022ilk,Kamran:2024xob,Foreman:2024kzw}. Power-spectrum measurements enable baryon acoustic oscillation studies~\citep{Rhook:2008gh,Ostergaard:2024brd}, growth-rate constraints, and inference of key astrophysical parameters such as the \HI\ bias~\citep{Padmanabhan:2014zma,Padmanabhan:2015wja,Sarkar:2018gcb,Sinigaglia:2020mut}. However, non-linear gravitational evolution and non-linear biasing generate phase correlations in the matter and \HI\ density fields that are invisible to two-point statistics~\citep{Bernardeau:2001qr,Sefusatti:2006pa}. These higher-order correlations are not a small correction: they encode both the non-linear mode coupling of gravity and the astrophysics of how \HI\ populates halos, and they become increasingly relevant on the quasi-linear scales that dominate the statistical power of next-generation surveys.

The bispectrum---the Fourier transform of the three-point correlation function---is the lowest-order statistic sensitive to these non-Gaussian correlations~\citep{Scoccimarro:1997st,Matarrese:1997sk,Scoccimarro:2000sn,Scoccimarro:2004tg,Sefusatti:2006pa,Gil-Marin:2014pva,Shaw:2021pgy,Mazumdar:2022ynd,Nandi:2024cib}. Its dependence on triangle configurations (e.g.\ equilateral, squeezed, and stretched) directly probes distinct mode-coupling channels and bias contributions, providing shape information that can break degeneracies that persist in power-spectrum analyses~\citep{Sarkar:2019ojl,Karagiannis:2020dpq,Ivanov:2021kcd,Ivanov:2023qzb,Gualdi:2020ymf,Gualdi:2021yvq}. This makes the bispectrum an especially attractive target for \texttt{SKA-MID}: the survey volume and sensitivity that enable high signal-to-noise power-spectrum measurements on quasi-linear scales also open the door to statistically significant three-point measurements~\citep{Karagiannis:2020dpq,Sarkar:2019ojl,Raste:2023zra}.

In redshift space, line-of-sight peculiar velocities imprint characteristic anisotropies through large-scale Kaiser distortions~\citep{Kaiser:1987qv,Hamilton:1997zq} and small-scale Fingers-of-God (FoG) damping~\citep{Jackson:1971sky,Scoccimarro:2004tg,Sarkar:2019nak}. For the bispectrum, a natural and information-preserving way to treat this anisotropy is to expand the signal in spherical harmonics with respect to the line-of-sight direction~\citep{Scoccimarro:1999ed,Sugiyama:2018yzo,Durrer:2020orn}. The resulting redshift-space bispectrum multipoles $B_\ell^m$ separate the isotropic component ($\ell=0$) from higher-order angular dependence~\citep{Gualdi:2020ymf}. Experience from galaxy surveys indicates that the lowest-order multipoles capture most of the redshift-space information, while higher-$\ell$ (where $\ell$ denotes the degree of anisotropy and $m$ labels the azimuthal dependence of the bispectrum with respect to the line-of-sight direction ) moments encode subtler angular structure that is progressively harder to measure at fixed sensitivity~\citep{Nan:2017oaq,Mazumdar:2022ynd,Pal:2025hpl}. A multipole-based approach therefore provides both a compact description of anisotropy and a practical pathway for forecasting and measurement.

Beyond their role in \HI\ astrophysics, higher-order statistics offer a powerful route to fundamental-physics tests, including constraints on the laws of gravity on cosmological scales~\citep{Koyama:2015vza,Alam:2020jdv}. Many modified-gravity (MG) theories predict scale-dependent changes to structure growth and altered non-linear mode coupling~\citep{Clifton:2011jh,Joyce:2014kja}. In scalar--tensor scenarios, a dynamical scalar field mediates an additional interaction and modifies the relation between density and gravitational potential~\cite{Horndeski:1974wa,Koyama:2013paa,Kobayashi:2019hrl}. The Hu--Sawicki $f(R)$ model~\citep{Hu:2007nk} is a well-studied example in which the effective gravitational strength is enhanced below the Compton wavelength of the scalar degree of freedom~\cite{DeFelice:2010aj,Sotiriou:2008rp}, leading to scale-dependent linear growth and modified higher-order perturbation kernels~\citep{Koyama:2009me,Taruya:2014faa,Bose:2016qun,Aviles:2017aor,Aviles:2018saf,Rodriguez-Meza:2023rga}. These effects can imprint distinctive, configuration-dependent signatures in the bispectrum and its multipoles~\citep{Yamauchi:2017ibz,Hirano:2018uar,Bose:2018zpk,Pal:2025zep}, thereby complementing power-spectrum constraints~\citep{Aviles:2024zlw,Rodriguez-Meza:2023rga,Berti:2022ilk}.

In this work, we develop a tree-level perturbation-theory framework for the post-reionization 21-cm bispectrum in redshift space, incorporating local Eulerian bias up to second order, large-scale Kaiser distortions, and a phenomenological description of FoG suppression. We apply the same modeling pipeline to both a fiducial $\Lambda$CDM cosmology and the Hu--Sawicki $f(R)$ MG model, consistently accounting for scale-dependent linear growth and modified second-order mode-coupling kernels. We then assess the prospects for measuring bispectrum multipoles with an \texttt{SKA-MID}-like IM survey~\citep{schediwy2019mid,MeerKLASS:2017vgf}. By incorporating instrumental noise, survey volume, and mode counting into the full bispectrum covariance, we forecast the signal-to-noise of the anisotropic multipoles and of MG-induced deviations relative to General Relativity. Our analysis is designed to clarify (i) which multipoles dominate the observable information, (ii) which triangle configurations maximize sensitivity, and (iii) where, in scale and redshift, \texttt{SKA-MID} is most capable of exploiting non-linear information beyond the power spectrum.

The paper is organized as follows. In Section~\ref{sec:bispectrum_formalism}, we present the modeling of the 21-cm bispectrum, including the real-space formulation, redshift-space distortions and multipole decomposition, and the extension to modified gravity. Section~\ref{sec:snr} describes the \texttt{SKA-MID} survey specifications, thermal-noise and covariance calculations, and the SNR forecasting methodology. In Section~\ref{subsec:snr_results}, we present our results for the bispectrum multipoles and the expected SNR as a function of triangle configuration, scale, and redshift, highlighting the configurations that maximize the detectability of MG signatures. Finally, Section~\ref{sec:conclusion} summarizes our findings and outlines directions for future work.

\section{Modeling the 21\texorpdfstring{\,}{ }cm bispectrum}
\label{sec:bispectrum_formalism}

\subsection{Real-space 21-cm field and bias model}
\label{subsec:real_space_formalism}

The 21-cm differential brightness temperature at redshift $z$ can be written as a sum of a mean component and fluctuations:
\begin{equation}
    T_b(\mathbf{x},z) = \overline{T}_b(z)\,\big[1 + \delta_{\text{H}\textsc{i}}(\mathbf{x},z)\big]~,
    \label{eq:Tb}
\end{equation}
where $\delta_{\text{H}\textsc{i}} = (\rho_{\text{H}\textsc{i}}-\overline{\rho}_{\text{H}\textsc{i}})/\overline{\rho}_{\text{H}\textsc{i}}$ is the \HI\ density contrast. The mean brightness temperature $\overline{T}_b$ is given by~\citep{Camera:2015yqa,Padmanabhan:2023hfr,Padmanabhan:2014zma,Castorina:2016bfm,Berti:2022ilk} :
\begin{equation}
    \overline{T}_b(z) \approx 0.566\,h \frac{H_0}{H(z)} \left(\frac{\Omega_{\text{H}\textsc{i}}(z)}{0.003}\right) (1+z)^2~{\rm mK},
    \label{eq:Tb_mean}
\end{equation}
where $H_0$ is the Hubble constant, $H(z)$ is the Hubble parameter at redshift $z$, and $\Omega_{\text{H}\textsc{i}}(z)$ is the \HI\ density fraction relative to the present critical density. Observations and simulations~\citep{Padmanabhan:2023hfr,Castorina:2016bfm,Villaescusa-Navarro:2014cma} indicate that $\Omega_{\text{H}\textsc{i}}(z)$ remains roughly constant for $0 \lesssim z \lesssim 6$ at a value $\Omega_{\text{H}\textsc{i}} \sim 4.7\times10^{-4}$.

On large scales in the post-reionization era, \HI\ can be treated as a biased tracer of the underlying matter density field. We adopt a local Eulerian bias expansion up to second order~\citep{Desjacques:2016bnm}:
\begin{equation}
    \delta_{\text{H}\textsc{i}}(\mathbf{x},z) = b_1(z)\,\delta_m(\mathbf{x},z) + \frac{b_2(z)}{2}\,\delta_m^2(\mathbf{x},z) + \cdots~,
    \label{eq:bias}
\end{equation}
where $\delta_m$ is the matter overdensity, and $b_1$, $b_2$ are the linear and quadratic \HI\ bias parameters, respectively. We use redshift-dependent bias values fitted from simulations of the real-space \HI\ bispectrum~\cite{Sarkar:2019ojl}:
\begin{eqnarray}
    b_1(z) &=& 0.653 + 0.252\,z + 0.0196\,z^2~, \label{eq:b1z}\\
    b_2(z) &=& -0.365 + 0.0121\,z^2 + 0.00217\,z^4~. \label{eq:b2z}
\end{eqnarray}
At $z= 2$, for example, these give $b_1 \sim 1.23$ and $b_2 \sim -0.28$. The linear \HI\ power spectrum is related to the matter power spectrum $P_m(k,z)$ by $P^r_{\text{H}\textsc{i}}(k,z) = b_1^2(z)\,P_m(k,z)$ (in units of $h^{-3}\,{\rm Mpc}^3$). Correspondingly, the 21-cm brightness-temperature power spectrum is $P_{21}(k,z) = \overline{T}_b^2(z)\,P^r_{\text{H}\textsc{i}}(k,z)$ (with units mK$^2\, h^{-3}\,{\rm Mpc}^3$). Here $k\equiv|\mathbf{k}|$ denotes the magnitude of the Fourier wavevector $\mathbf{k}$ (i.e.\ the comoving wavenumber), with units of $h\,\mathrm{Mpc}^{-1}$.

At tree level in perturbation theory, the real-space bispectrum of the \HI\ density field (ignoring redshift-space effects) is given by the sum of contributions from non-linear gravity (mode coupling) and non-linear bias~\citep{Sarkar:2019nak}:
\begin{align}\label{eq:B_real}
B_{\text{H}\textsc{i}}(\mathbf{k}_1,\mathbf{k}_2,\mathbf{k}_3,z) &= \frac{1}{b_1(z)} \Big[ 2F_2(\mathbf{k}_1,\mathbf{k}_2)P^r_{\text{H}\textsc{i}}(k_1,z)P^r_{\text{H}\textsc{i}}(k_2,z) \nonumber  \\&    \hspace{-2.3em}  + \text{cyc.} \Big] + \frac{b_2(z)}{{b_1}^2(z)} \Big[ P^r_{\text{H}\textsc{i}}(k_1,z)P^r_{\text{H}\textsc{i}}(k_2,z) + \text{cyc.} \Big],
\end{align}
where $(k_1,k_2,k_3)$ form a closed triangle ($\mathbf{k}_1+\mathbf{k}_2+\mathbf{k}_3=0$) and ``cyc.'' denotes cyclic permutations of $(k_1,k_2,k_3)$. The first term in Eq.~\eqref{eq:B_real} arises from non-linear gravitational clustering (with $F_2$ being the second-order density perturbation kernel), while the second term arises from the quadratic bias $b_2$. We adopt the Einstein--de Sitter (EdS) approximation for the standard perturbation theory kernels, which is accurate to within a few percent at $1\leq z \leq 6$. In an EdS universe, the symmetrized second-order density kernel and velocity-divergence kernel are~\citep{Scoccimarro:2000sn,Bernardeau:2001qr}:
\begin{eqnarray}
    F_2(k_1,k_2, \mu) = \frac{5}{7} - \frac{1}{2}\left(\frac{k_1}{k_2} + \frac{k_2}{k_1}\right)\mu + \frac{2}{7}\mu^2~, \label{eq:F2EdS}\\
    G_2(k_1,k_2, \mu) =
\frac{3}{7} - \frac{1}{2}\left(\frac{k_1}{k_2}+\frac{k_2}{k_1}\right)\mu
+ \frac{4}{7}\mu^2~, \label{eq:G2EdS}
\end{eqnarray}
with $\mu = -\hat{\mathbf{k}}_1 \cdot \hat{\mathbf{k}}_2$. 
Note that, the velocity-divergence kernel $G_2$ does not enter the real-space bispectrum expression; we introduce it here purely for convenience and will make use of it in the next section.
The distinct functional dependence of the two terms in Eq.~\eqref{eq:B_real}---the $F_2$ term depends on triangle shape (angles between $\mathbf{k}_1$ and $\mathbf{k}_2$) while the $b_2$ term depends only on magnitudes---allows in principle for simultaneous determination of $b_1$ and $b_2$ from bispectrum measurements, complementing the power spectrum.

\subsection{Redshift-space distortions and bispectrum multipoles}
\label{sec:HI_bispectrum_rsd}

The observed 21-cm signal is affected by redshift-space distortions (RSD) because peculiar velocities cause the apparent position (in redshift/frequency space) of each \HI\ emitter to shift along the line of sight. In linear theory, this is encapsulated by the Kaiser formula \citep{Kaiser:1984sw}: in Fourier space, the \HI\ overdensity in redshift space is $\delta^s_{\text{H}\textsc{i}}(\mathbf{k}_i) = (b_1 + f\mu_i^2)\,\delta_m(\mathbf{k}_i)$, where $f(z) \equiv d\ln D/d\ln a$ is the linear growth rate and $\mu_i \equiv \cos\theta_i$ is the cosine of the angle between $\mathbf{k}_i$ and the line-of-sight direction. The factor $(b_1 + f\mu_i^2)$, often denoted $Z_1(\mathbf{k}_i)$, enhances clustering for modes with a line-of-sight component (since infall velocities increase the apparent contrast along $\mathbf{k}_{i,\parallel}$, where 
$\mathbf{k}_{i,\parallel}\equiv(\mathbf{k}_i\cdot\hat{\mathbf{n}})\,\hat{\mathbf{n}}$ is the component of $\mathbf{k}_i$ parallel to the line of sight $\hat{\mathbf{n}}$). On small scales, the random virial motions of \HI\ inside halos lead to FoG damping of the power spectrum. We model this in power spectrum with a dispersion term. In particular, we assume a Gaussian form for the FoG suppression~\citep{Peebles:1980yev,Hikage:2015wfa,BaleatoLizancos:2025wdg}: 
\begin{equation}
    \mathcal{D}^{P}_{\rm FoG}(k_i,\mu_i) = \exp\left[-\frac{1}{2}(k_i\,\mu_i\,\sigma_p)^2\right]~, \label{eq:FoG_power}
\end{equation}
where $\sigma_p$ is the one-dimensional velocity dispersion of \HI\ (in $h^{-1}{\rm Mpc}$). 
Combining the Kaiser enhancement and FoG suppression, the redshift space \HI\ power spectrum is given by, 
\begin{equation}
P_{\text{H}\textsc{i}}^s(k, z) = \bar{T_b}^2(z) \, \mathcal{D}_{\rm FOG}^P(k, z) \, Z_1^2(k, z) \, P_{\text{H}\textsc{i}}^r(k, z),
\end{equation}

Including these RSD effects, the tree-level redshift-space \HI\ bispectrum can be written in the plane-parallel approximation as:
\begin{align}\label{eq:B_rsd}
    B^{s}_{\text{H}\textsc{i}}(\mathbf{k_1},\mathbf{k_2},\mathbf{k_3}; z) &= \overline{T}_b^3(z)\,\mathcal{D}^{B}_{\rm FoG}(k_1,k_2,k_3,\{\mu_i\}) \Big[2\,Z_1(\mathbf{k}_1) \nonumber \\& \hspace{-3.0em} \times  Z_1(\mathbf{k}_2)Z_2(\mathbf{k}_1,\mathbf{k}_2)\,P^{r}_{\text{H}\textsc{i}}(k_1)P^{r}_{\text{H}\textsc{i}}(k_2) + \text{2 perm.}\Big]~, 
\end{align}
where $P^r_{\text{H}\textsc{i}}$ is the real-space \HI\ power spectrum, $Z_1(\mathbf{k}) = b_1 + f\mu^2$ as above, and $Z_2(\mathbf{k}_i,\mathbf{k}_j)$ is the second-order redshift-space kernel:
\begin{align}
    Z_2(\mathbf{k}_i,\mathbf{k}_j) &= b_1\,F_2(\mathbf{k}_i,\mathbf{k}_j) + f\,\mu_{ij}^2\,G_2(\mathbf{k}_i,\mathbf{k}_j) + \frac{b_2}{2} \nonumber \\& \hspace{1.0em} + \frac{f\,\mu_{ij}}{2}\Big[\mu_i \frac{k_i}{k_j} Z_1(\mathbf{k}_j) + \mu_j \frac{k_j}{k_i} Z_1(\mathbf{k}_i)\Big]~. 
    \label{eq:Z2kernel}
\end{align}
Here $\mu_{ij}$ denotes the cosine of the angle between $\mathbf{k}_i + \mathbf{k}_j$ and the line of sight. The factor $\mathcal{D}^B_{\rm FoG}(k_1,k_2,k_3)$ represents FoG damping of the bispectrum, which we take as 
\begin{equation}
    \mathcal{D}^{B}_{\rm FoG}(k_1,k_2,k_3,\{\mu_i\}) = \exp\left[-\frac{1}{2}\sigma_p^2 (k_1^2\mu_1^2 + k_2^2\mu_2^2 + k_3^2\mu_3^2)\right]~,
    \label{eq:FoG_bispec}
\end{equation}
the natural generalization of Eq.~\ref{eq:FoG_power} to three-wavevector configurations. In Eq.~\ref{eq:Z2kernel}, the terms proportional to $b_1$ and $b_2$ correspond to density contributions, whereas terms with $f$ involve velocity contributions. The last term (in brackets) arises from one linear RSD factor $Z_1$ times one nonlinear velocity coupling; it depends on the angle of each $\mathbf{k}_i$ via $\mu_i = \cos\theta_i$ (the cosine of the angle between $\mathbf{k}_i$ and the line of sight).
In this work, we adopt $\sigma_p(z)$ from Table~1 of \citet{Sarkar:2018gcb}. We apply the same dispersion $\sigma_p$ in both the power spectrum and bispectrum FoG factors.

It is convenient to parameterize the configuration of the closed triangle, $\mathbf{k}_1 + \mathbf{k}_2 + \mathbf{k}_3 = \mathbf{0}$, by invariants rather than the three $\mathbf{k}_i$ vectors. We follow earlier works \citep{Bharadwaj:2020wkc, Mazumdar:2020bkm, Pal:2025hpl} by specifying a triangle with $(k_1,\mu,t)$, where $k_1$ is the length of the largest wave vector, $t = k_2/k_1$ (ratio of the second largest wave vector to the largest), and $\mu = -\hat{\mathbf{k}}_1 \cdot \hat{\mathbf{k}}_2$ is the cosine of the angle between $\mathbf{k}_1$ and $\mathbf{k}_2$. For a valid triangle, $0.5 \le t \le 1$ and $0.5 \le \mu \le 1$, with the third side $k_3$ determined by $k_3 = k_1\sqrt{1 - 2\mu\,t + t^2}$ and requiring $2\mu t \ge 1$. Some notable configurations in the $(\mu,t)$ plane are: equilateral triangles $(t=1,\mu=0.5)$, squeezed triangles $(\mu \to 1,\,t \to 1)$ where $\mathbf{k_1}\approx -\mathbf{k_2}$ and $\mathbf{k_3}\approx0$, 
linear triangles $(\mu\to1)$ where the $k$-vectors, $\mathbf{k_1},-\mathbf{k_2},\mathbf{-k_3}$, are colinear, 
stretched triangles $(\mu \to 1,\,t \to 0.5)$ where $\mathbf{k_2}=\mathbf{k_3} \approx -\mathbf{k_1}/2$, right-angle triangles $(\mu = t)$, and two special isosceles cases: $t=1$ (\emph{L-isosceles}, with the two long sides equal) and $2\mu t=1$ (\emph{S-isosceles}, with the two short sides equal). We will visualize results on the $(\mu,t)$ plane to understand how different shapes affect the bispectrum.

Using the bispectrum model defined above, the bispectrum can be formally decomposed by integrating over the orientation of the triangle in space. This multipole decomposition formalism was introduced in \citet{Bharadwaj:2020wkc} and we summarize it briefly here for completeness.  
The reference triangle is initially defined in the $x$-$z$ plane, with the $z$-axis aligned along the line of sight, so that the wave vectors are given by
\begin{equation}
\mathbf{k}_1 = k_1 \hat{z}, \, \,\,\, \mathbf{k}_2 = k_1 t(-\mu \, \hat{z} + \sqrt{1-\mu^2} \,\hat{x}), \, \,\,\, \mathbf{k}_3 = -(\mathbf{k}_1 + \mathbf{k}_2).
\end{equation}
To describe arbitrary orientations of the triangle with respect to the line of sight, we can rotate this configuration using Euler angles $(\alpha,\beta,\gamma)$, keeping the line-of-sight direction fixed as $\hat{\mathbf{n}}=\hat{z}$.
The resulting direction cosines between each wavevector and the line of sight are
\begin{align}
\mu_1 = \cos \beta = p_z, \quad \mu_2 = -\mu p_z - \sqrt{1-\mu^2} p_x, \quad \nonumber \\\mu_3 = -\frac{k_1 \mu_1 + k_2 \mu_2}{k_3} = -(1-t\mu)p_z + \frac{t \sqrt{1-\mu^2}}{s} p_x,
\end{align}
where $\mathbf{p}=(p_x,p_y,p_z)$ specifies the orientation of the triangle. 
The redshift-space \HI\ bispectrum can be expanded in spherical harmonics with respect to $\hat{\mathbf{p}}$ as
\begin{equation}
B_{\text{H}\textsc{i}}^s(k_1, \mu, t, \hat{\mathbf{p}}) = \sum_{\ell=0}^{\infty} \sum_{m=-\ell}^{\ell} B_{\ell}^m(k_1, \mu,t ) \, Y_{\ell}^m(\hat{\mathbf{p}}),
\end{equation}
where $B_{\ell}^m(k_1, \mu,t)$ are the bispectrum multipoles, capturing the anisotropic structure induced by redshift-space distortions.
The multipoles are obtained via the standard projection:
\begin{equation}\label{eq:blm}
B_{\ell}^m(k_1, \mu,t) = \int d\Omega_{\hat{\mathbf{p}}} \, B_{\text{H}\textsc{i}}^s(k_1, \mu,t,\hat{p}) \, {[Y_{\ell }^{m}(\hat{\mathbf{p}})]}^*.
\end{equation}

\subsection{Modified gravity: Hu--Sawicki \texorpdfstring{$f(R)$}{f(R)} model}
\label{seubsec:beyond_GR}
The results presented so far are true for General Relativity (GR).
In order to investigate the beyond GR effects in the post-reionization 21-cm bispectrum, we consider the Hu-Sawicki (HS) $f(R)$ gravity model~\citep{Hu:2007nk,DeFelice:2010aj}. In this theory, the Einstein--Hilbert action is modified by replacing $R$ with $R+f(R)$. The HS functional form is designed to yield late-time cosmic acceleration while remaining consistent with the $\Lambda$CDM expansion history \citep{Lobo:2014ara,Linder:2010py,Turner:2007qg}.

In the quasi-static regime relevant for large-scale structure, the model predicts a scale-dependent enhancement of the gravitational force mediated by an additional scalar degree of freedom. This enhancement becomes relevant on scales shorter than the Compton wavelength associated with the scalar field, and is dynamically suppressed in high-density environments via the chameleon screening mechanism \citep{Khoury:2003aq,Li:2011qda}.

A convenient perturbative description is obtained in a scalar--tensor formulation in the Newtonian gauge, introducing the scalar perturbation $\delta f_R \equiv f_R - f_{R0}$ sourced by matter density fluctuations $\delta\rho = \bar{\rho}\,\delta$. Here $\bar{\rho}$ denotes the background density, while $R_0$ denotes the background curvature. The modified Poisson equation in Fourier space can be written as \citep{Aviles:2023fqx,Rodriguez-Meza:2023rga}
\begin{equation}
    -\frac{k^2}{a^2}\,\Phi \;=\; 4\pi G\,\bar{\rho}\,\delta\,\tilde{\mu}(k,a) \;+\; \mathcal{S}(k),
\end{equation}
where
\begin{align}
\tilde{\mu}(k,a) &= 1 + \frac{1}{3}\,\frac{k^2}{k^2 + a^2 m^2(a)}, \\
\mathcal{S}(k) &= -\frac{1}{6}\,\frac{k^2}{k^2 + a^2 m^2(a)}\,\delta I.
\end{align}
The function $\tilde{\mu}(k,a)$ parameterizes the enhancement of the effective gravitational strength: on scales $k \gg a m$ gravity is strengthened by up to a factor of $4/3$, while GR is recovered on large scales $k \ll a m$. The term $\mathcal{S}(k)$ encapsulates non-linear self-interactions of the scalar field through $\delta I$, which can be expanded perturbatively as \citep{Aviles:2023fqx,Rodriguez-Meza:2023rga}
\begin{align}
\delta I(\delta f_R) &= \frac{1}{2} \int_{k_1 + k_2 = k} M_2(k_1, k_2)\, \delta f_R(k_1)\delta f_R(k_2) \nonumber \\
& + \frac{1}{6} \int_{\sum k_i = k} M_3(k_1, k_2, k_3) \delta f_R(k_1)\delta f_R(k_2)\delta f_R(k_3)\nonumber  \\ &
+ \cdots,
\end{align}
with $M_n \equiv \left.\dfrac{d^n R}{d f_R^n}\right|_{f_{R0}}$. For the HS model with $n=1$, these functions depend only on time through background quantities, and are given by
\begin{align}
M_1(a) &= \frac{3}{2}H_0^2 \frac{\big(\Omega_{m,0}a^{-3}+4\Omega_{\Lambda,0}\big)^3}{|f_{R0}|\big(\Omega_{m,0}+4\Omega_{\Lambda,0}\big)^2}, \\
M_2(a) &= \frac{9}{4}H_0^2 \frac{\big(\Omega_{m,0}a^{-3}+4\Omega_{\Lambda,0}\big)^5}{|f_{R0}|^2\big(\Omega_{m,0}+4\Omega_{\Lambda,0}\big)^4}.
\label{eq:M2}
\end{align}

At linear order, the modification corresponds to a Yukawa-like fifth force with range set by the Compton wavelength, $\lambda_C \sim m^{-1}(a) \propto |f_{R0}|^{1/2}$; smaller values of $|f_{R0}|$ therefore correspond to weaker deviations from GR. These changes in the evolution of the gravitational potential lead to a scale-dependent growth of matter perturbations. This is captured by generalizing the linear evolution equation for the matter overdensity, $\delta^{(1)}(\mathbf{k},a)$, in terms of a scale-dependent growth factor $D_+(k,a)$ satisfying
\begin{equation}
\big[\mathcal{T} - A(k,a)\big]\,D_+(k,a) = 0,
\label{eq:Dplus_evolution}
\end{equation}
where $\mathcal{T}=\partial_t^2 + 2H\partial_t$ and
\begin{equation}
A(k,a) = \frac{3}{2}\Omega_m H^2\left(1 + \frac{2\beta^2 k^2}{k^2 + m^2 a^2}\right).
\label{eq:A_def}
\end{equation}
Here $\beta(a)$ denotes the universal coupling to matter (with $\beta^2=1/3$ in the HS-$f(R)$ model).
For computational consistency and for comparison across models, it is useful to normalize $D_+$ by the EdS solution at an early epoch $t_{\rm ini}$, where MG effects are negligible and $D_+(k,t_{\rm ini})\propto a(t_{\rm ini})$. The evolved linear power spectrum in a given MG theory then reads
\begin{equation}
P_L(k,t) = \left[\frac{D_+(k,t)}{D_+^{\Lambda \mathrm{CDM}}(k,t_0)}\right]^2 P_L^{\Lambda \mathrm{CDM}}(k,t_0),
\label{eq:PL_mg}
\end{equation}
where $t_0$ is a reference time, typically today.

While $D_+$ captures linear growth, describing non-linear structure formation requires higher-order perturbative corrections. At second order in the density contrast and velocity divergence, the non-linear evolution is governed by scale- and time-dependent kernels~\citep{Rodriguez-Meza:2023rga,Aviles:2023fqx},
\begin{align}
F_2(\mathbf{k}_1,\mathbf{k}_2,t) &= \frac{1}{2} + \frac{3}{14} \mathcal{A}(\mathbf{k}_1,\mathbf{k}_2,t) + \frac{\hat{\mathbf{k}}_1 \cdot \hat{\mathbf{k}}_2}{2}\left(\frac{k_1}{k_2} + \frac{k_2}{k_1}\right) \nonumber \\
&\quad + (\hat{\mathbf{k}}_1 \cdot \hat{\mathbf{k}}_2)^2\left[\frac{1}{2} - \frac{3}{14} \mathcal{B}(\mathbf{k}_1,\mathbf{k}_2,t) \right]. \label{eq:F2_HS}\\
G_2(\mathbf{k}_1,\mathbf{k}_2,t) &= \frac{3 \left[f(k_1) + f(k_2)\right]\mathcal{A}+3 \dot{\mathcal{A}}/H}{14 f_0}  \nonumber \\
&\quad + \frac{\hat{\mathbf{k}}_1 \cdot \hat{\mathbf{k}}_2}{2} \left[ \frac{f(k_2)}{f_0} \frac{k_2}{k_1} + \frac{f(k_1)}{f_0} \frac{k_1}{k_2} \right] \nonumber \\
&\quad + \left( \hat{\mathbf{k}}_1 \cdot \hat{\mathbf{k}}_2 \right)^2 \Big[ \frac{f(k_1) + f(k_2)}{2 f_0} \nonumber \\
&\quad  - \frac{3 \left[f(k_1) + f(k_2)\right] \mathcal{B} + 3 \dot{\mathcal{B}}/H}{14 f_0}.
\label{eq:G2_HS}
\end{align}
The functions $\mathcal{A}$ and $\mathcal{B}$ quantify deviations from EdS evolution and are defined by
\begin{align}
\mathcal{A} &= \frac{7\,D^{(2)}_{\mathcal{A}}(\mathbf{k}_1,\mathbf{k}_2)}{3\,D_+(k_1)\,D_+(k_2)}, \qquad
\mathcal{B} = \frac{7\,D^{(2)}_{\mathcal{B}}(\mathbf{k}_1,\mathbf{k}_2)}{3\,D_+(k_1)\,D_+(k_2)}.
\end{align}
In $\Lambda$CDM, $\mathcal{A}\simeq \mathcal{B}\simeq 1$ today.

The second-order growth functions $D^{(2)}_{\mathcal{A}}$ and $D^{(2)}_{\mathcal{B}}$ solve Green's-type equations sourced by the modified gravitational dynamics,
\begin{align}
D^{(2)}_{\mathcal{A}} &= \left(\mathcal{T} - A(k)\right)^{-1} \Big[ A(k) + (A(k) - A(k_1)) \frac{\mathbf{k}_1\cdot\mathbf{k}_2}{k_2^2} \nonumber \\
& \hspace{-2.0em} + (A(k) - A(k_2)) \frac{\mathbf{k}_1\cdot\mathbf{k}_2}{k_1^2} - S_2(\mathbf{k}_1,\mathbf{k}_2) \Big] D_+(k_1) D_+(k_2),
\label{eq:DA}\\
D^{(2)}_{\mathcal{B}} &= \left(\mathcal{T} - A(k)\right)^{-1} \left[ A(k_1) + A(k_2) - A(k) \right] D_+(k_1) D_+(k_2).
\label{eq:DB}
\end{align}
where, for brevity, $A(k)\equiv A(k,a)$ is understood at the relevant time.
The source term $S_2$ captures non-linear modifications due to scalar interactions and plays a central role in screening. For HS-$f(R)$, it takes the form \citep{Aviles:2023fqx,Rodriguez-Meza:2023rga}
\begin{equation}
S_2(\mathbf{k}_1,\mathbf{k}_2) =
\frac{36\,\Omega_m^2 H^4 \beta^6 a^4\,M_2(a)\,k^2}
{\big(k^2 + m^2 a^2\big)\big(k_1^2 + m^2 a^2\big)\big(k_2^2 + m^2 a^2\big)},
\label{eq:S2_fR}
\end{equation}
where $M_2$ is the model-dependent non-linear coupling defined in Eq.~\eqref{eq:M2}. Note that to compute the kernels up to second order, we only require the source correction controlled by $M_2$.

As discussed in the previous section, these second-order corrections enter the tree-level matter bispectrum, which encapsulates the non-Gaussianity induced by gravitational evolution. For detailed calculations of the second- and higher-order kernels in this framework, see Refs.~\cite{Aviles:2017aor,Aviles:2018qot,Aviles:2018saf,Aviles:2020cax,Aviles:2020wme,Aviles:2023fqx,Rodriguez-Meza:2023rga}.

\subsection{Multipoles of the 21-cm bispectrum in $\Lambda$CDM}
\label{subsec:multipoles}

\begin{figure*}
\centering
\subfloat[]{\includegraphics[width=0.32\textwidth]{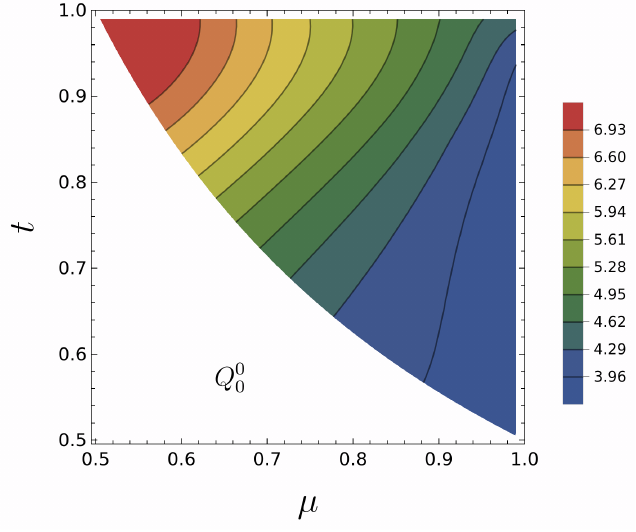}}
\subfloat[]{\includegraphics[width=0.32\textwidth]{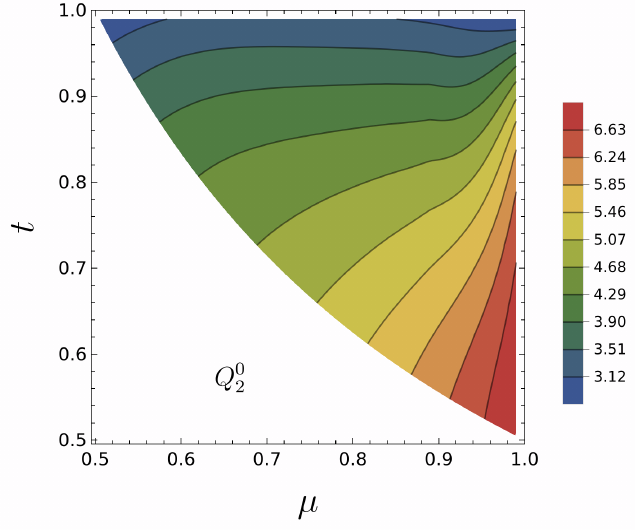}}
\subfloat[]{\includegraphics[width=0.32\textwidth]{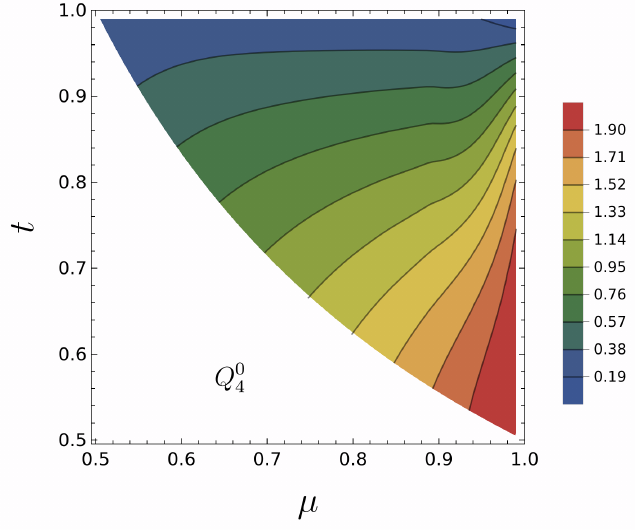}}\\

\subfloat[]{\includegraphics[width=0.32\textwidth]{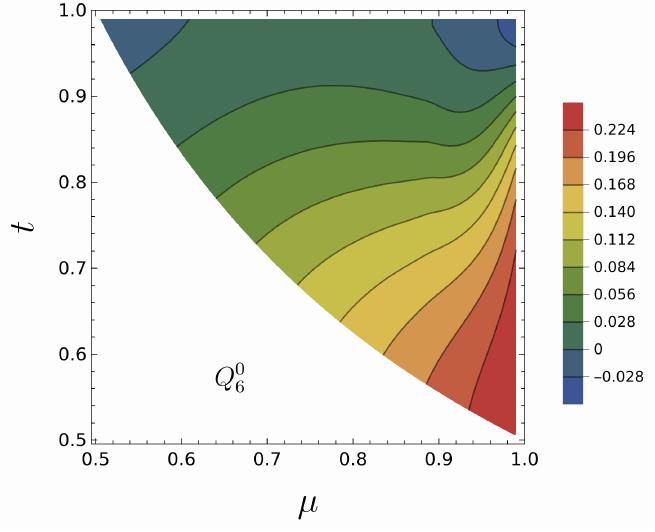}}
\subfloat[]{\includegraphics[width=0.32\textwidth]{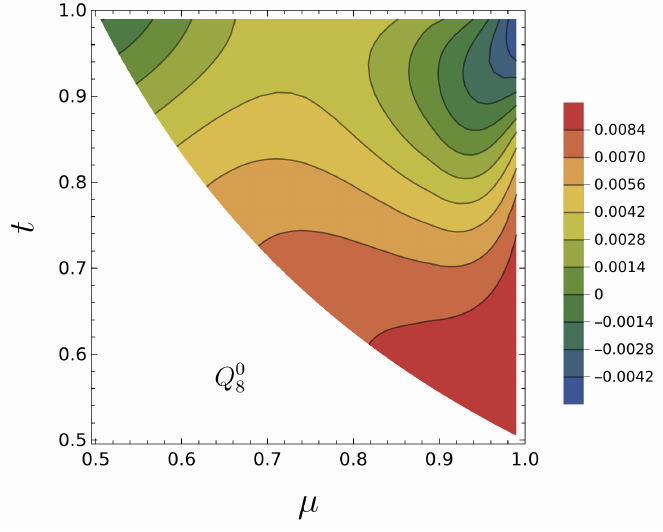}}
\caption{\justifying Reduced redshift-space \HI\ bispectrum multipoles $Q_\ell^0(\mu,t)$ in the $(\mu,t)$ plane for a fiducial $\Lambda$CDM cosmology, evaluated at $k_1=0.1\,h\,\mathrm{Mpc}^{-1}$ and $z=2$. The reduced multipoles are defined as $Q_\ell^m \equiv B_\ell^m/B_{\text{H}\textsc{i}}^{\rm r}$, where $B_\ell^m$ are the spherical-harmonic multipoles of the redshift-space bispectrum (including Kaiser RSD and FoG damping), and $B_{\text{H}\textsc{i}}^{\rm r}$ is the real-space bispectrum.
The monopole $Q_0^0$ shows a strong overall enhancement relative to real space across most of the allowed triangle domain, while the quadrupole $Q_2^0$ captures the leading anisotropic response and is largest toward nearly linear configurations ($\mu\to 1$), particularly for stretched triangles ($t\lesssim 0.8$). Higher-order $m=0$ multipoles ($\ell\ge 4$) are progressively suppressed and exhibit oscillatory structure across the $(\mu,t)$ plane, indicating that the dominant RSD-induced information is concentrated in the lowest multipoles.}
\label{fig:Qm0_multipoles}
\end{figure*}

\begin{figure*}
\centering
\subfloat[]{\includegraphics[width=0.25\textwidth]{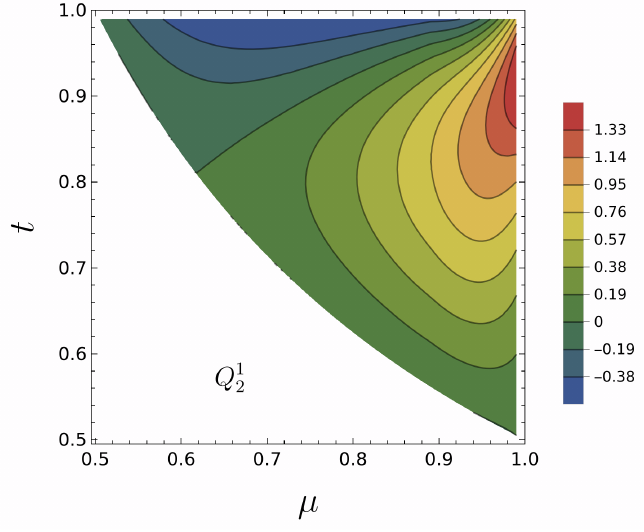}}
\subfloat[]{\includegraphics[width=0.25\textwidth]{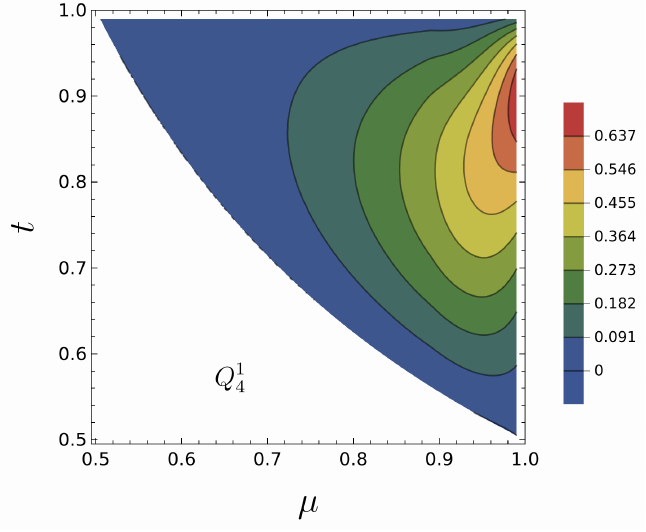}}
\subfloat[]{\includegraphics[width=0.25\textwidth]{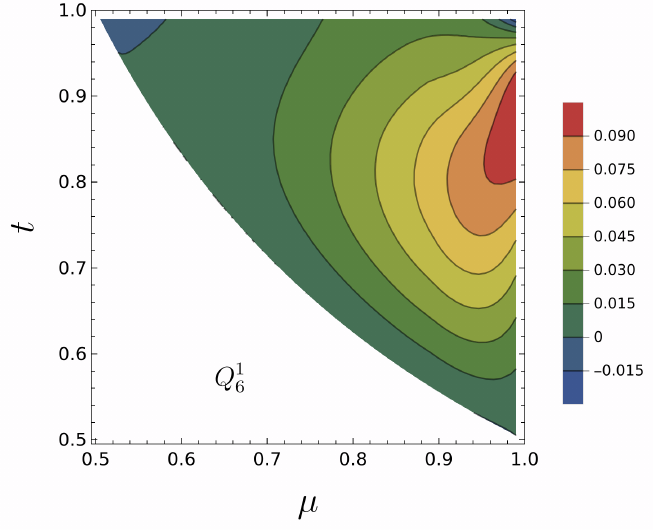}}
\subfloat[]{\includegraphics[width=0.25\textwidth]{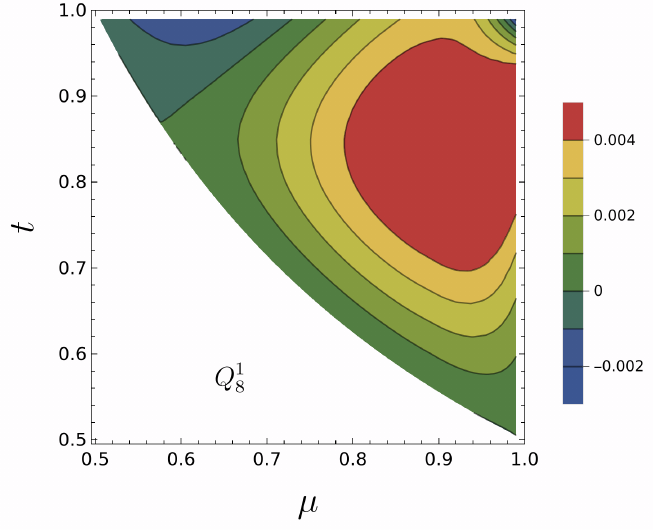}}\\

\subfloat[]{\includegraphics[width=0.25\textwidth]{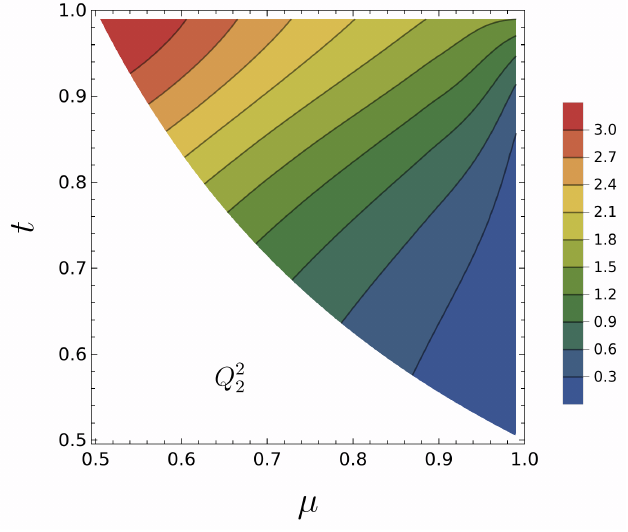}}
\subfloat[]{\includegraphics[width=0.25\textwidth]{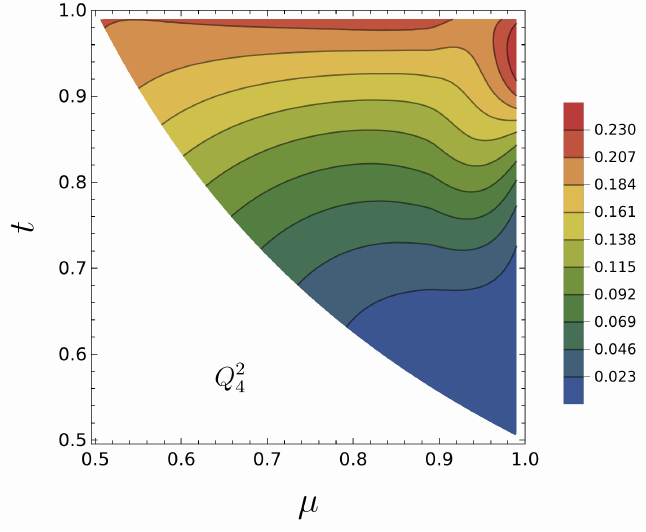}}
\subfloat[]{\includegraphics[width=0.25\textwidth]{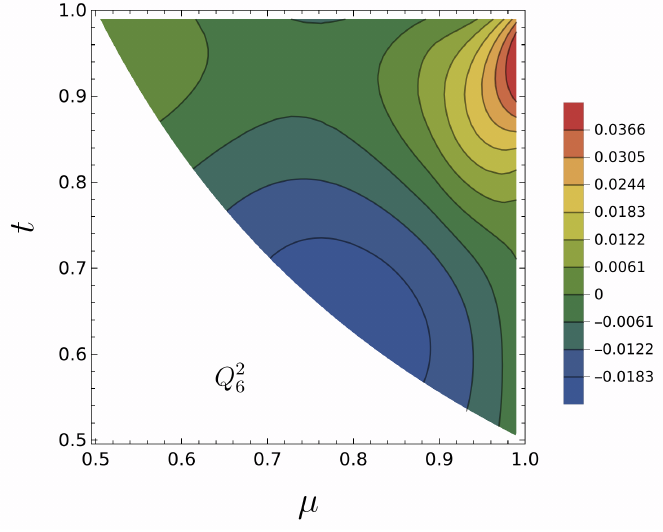}}
\subfloat[]{\includegraphics[width=0.25\textwidth]{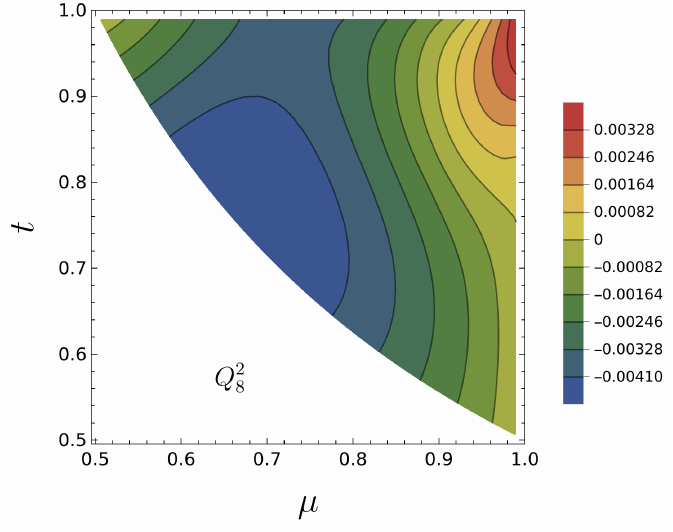}}
\caption{\justifying Reduced redshift-space \HI\ bispectrum multipoles $Q_\ell^m(\mu,t)$ in $\Lambda$CDM at $k_1=0.1\,h\,\mathrm{Mpc}^{-1}$ and $z=2$, shown for azimuthal modes $m=1$ (top row) and $m=2$ (bottom row) and for $\ell=2,4,6,8$ (left to right). For $m=1$, the dominant contribution is $Q_2^1$, with its largest amplitude localized near the squeezed/near-linear boundary of configuration space, while higher-$\ell$ modes are strongly suppressed by angular cancellations. For $m=2$, the leading signal arises from $Q_2^2$, with $\ell\ge 4$ multipoles exhibiting weaker, more oscillatory patterns and substantially smaller amplitudes. Overall, these panels illustrate that measurable azimuthal anisotropy is primarily carried by the lowest-order multipoles, motivating the focus on low $(\ell,m)$ in subsequent detectability forecasts.}

\label{fig:Qm1m2_multipoles}
\end{figure*}

In this section, we quantify how redshift-space distortions (RSD) reshape the \HI\ bispectrum by comparing its redshift-space multipoles to the corresponding real-space bispectrum. We work with the reduced multipoles
\begin{equation}
Q_\ell^m(k_1,\mu,t)\equiv \frac{B_\ell^m(k_1,\mu,t)}{B_{\text{H}\textsc{i}}^{\rm r}(k_1,\mu,t)}\,,
\end{equation}
which isolate anisotropic RSD effects while factoring out the overall real-space configuration dependence.\footnote{Note that our convention differs from the commonly used ``reduced bispectrum'' definition $Q_\ell^m = B_\ell^m/P^2(k_1)$, where $P(k_1)$ is the power spectrum.}
Here $B_{\text{H}\textsc{i}}^{\rm r}$ is the real-space \HI\ bispectrum defined in Eq.~\ref{eq:B_real}, while $B_\ell^m$ are the spherical-harmonic multipoles of the tree-level redshift-space \HI\ bispectrum defined in Eq.~\ref{eq:B_rsd}. We compute the linear matter power spectrum using \texttt{CLASS} \citep{Lesgourgues:2011re,Blas:2011rf}, fixing the cosmological parameters to
$\omega_{\rm cdm}=0.1201$, $\omega_{\rm b}=0.02238$, $h=0.6781$, $A_s=2.1\times 10^{-9}$, $n_s=0.9660$, and $\tau_{\rm reio}=0.0543$.
In general we find non-zero contributions up to $\ell\le 8$ and $m\le 6$; in the main text we focus on $m=0,1,2$ and defer higher-$m$ results to the appendix.
Unless stated otherwise, we present results at a fiducial redshift $z=2$, which lies well within the SKA1-MID Band~1 frequency coverage for the redshifted 21-cm line.

Figs.~\ref{fig:Qm0_multipoles} and \ref{fig:Qm1m2_multipoles} show $Q_\ell^m$ in the $(\mu,t)$ plane at $k_1=0.1\,h\,{\rm Mpc}^{-1}$ and $z=2$, illustrating how RSD imprint configuration-dependent anisotropies across the full allowed triangle space. We first discuss the isotropic family ($m=0$), and then the leading azimuthal modes ($m=1,2$).

\paragraph*{Isotropic family ($m=0$):}
The monopole $(\ell=0,m=0)$ captures the orientation-averaged redshift-space enhancement. As shown in the left panel of the top row of Fig.~\ref{fig:Qm0_multipoles}, the monopole is strongly enhanced over most of the $(\mu,t)$ plane, with $Q_0^0>1$ nearly everywhere. The enhancement is most pronounced near the equilateral configuration $(\mu\to 0.5,\,t\to 1)$, where $B_0^0\simeq 7\times B_{\text{H}\textsc{i}}^{\rm r}$. Toward the linear-triangle limit the amplification decreases, while stretched and squeezed configurations typically show an enhancement of order $\sim 4$ relative to real space. The monopole also exhibits appreciable redshift dependence, which we do not show explicitly here but note for completeness. For example, at $z=1$ the location of the maximum enhancement shifts toward stretched configurations, and equilateral triangles become highly suppressed. Overall, this behavior reflects the sensitivity of $Q_0^0$ to coherent line-of-sight motions that boost the bispectrum amplitude in redshift space.

The quadrupole $(\ell=2)$ encodes the leading anisotropic response to RSD. The $m=0$ mode is largest toward $\mu\to 1$, particularly for stretched configurations with $t\lesssim 0.8$, and gradually weakens toward the L-isosceles and equilateral limits. In contrast to the monopole, the shape of the $\ell=2$ pattern shows relatively mild redshift evolution, aside from an overall change in amplitude. The $\ell=4$ contribution remains non-negligible in parts of configuration space: stretched triangles reach $\sim 2\times B_{\text{H}\textsc{i}}^{\rm r}$, whereas in the equilateral and L-isosceles limits the signal is suppressed to $\sim 0.1\times B_{\text{H}\textsc{i}}^{\rm r}$. Finally, the $\ell=6$ and $\ell=8$ isotropic components are strongly suppressed and display alternating regions of enhancement and suppression across the $(\mu,t)$ plane, reflecting the oscillatory nature of higher-order angular projections. Taken together, these results indicate that the dominant RSD-induced information is concentrated in the lowest multipoles, with higher $\ell$ providing subdominant but structured corrections.

\paragraph*{Azimuthal modes ($m=1,2$):}
Fig.~\ref{fig:Qm1m2_multipoles} shows the leading azimuthal dependence of $Q_\ell^m$ for $m=1$ (top) and $m=2$ (bottom), for $\ell=2,4,6,8$.

For $m=1$, the dominant contribution is $Q_2^1$, whose amplitude ranges approximately from $-0.4$ to $1.3$, with a maximum localized near the squeezed, with $(\mu,t)\approx(1,0.9)$. This reflects the fact that $m=1$ isolates a dipolar line-of-sight coupling that is strongest when two sides of the triangle are closely aligned. The next multipole, $Q_4^1$, shows a qualitatively similar but more localized pattern with a reduced amplitude of order $\sim 0.6$, consistent with stronger angular cancellations at higher $\ell$. The higher modes $Q_6^1$ and $Q_8^1$ are further suppressed by two to three orders of magnitude (typical values in the range $10^{-3}$--$10^{-2}$) and exhibit alternating positive and negative patches across configuration space.

For $m=2$, the dominant signal arises from $Q_2^2$, which reaches values from $\sim 0.3$ up to nearly $3$. At $z=2$, its configuration dependence closely resembles that of the monopole, although it remains redshift sensitive; in particular, the enhancement can become most prominent near stretched configurations at different redshifts. The $\ell=4$ mode $Q_4^2$ shows its largest relative enhancement in squeezed and L-isosceles configurations, but with a comparatively small amplitude: in the region $t\ge 0.8$ and $0.5\le \mu \le 1$ it is only $\sim 0.2\times B_{\text{H}\textsc{i}}^{\rm r}$, while stretched configurations are suppressed by about two orders of magnitude relative to squeezed triangles. As in the $m=1$ case, the $\ell=6$ and $\ell=8$ modes follow a similar qualitative pattern (squeezed configurations positive and stretched negative) but are strongly suppressed, at the level of $10^{-2}$--$10^{-3}$.

Across all $m=0,1,2$ modes shown here, the monopole and quadrupole provide the largest and most robust contributions, while higher multipoles are increasingly diminished by angular cancellations and by non-linear velocity effects such as FoG damping. Motivated by this hierarchy, we restrict the observational analysis in later sections to the lowest-order modes, focusing in particular on multipoles up to $Q_4^1$.
\subsection{Multipoles of the 21-cm bispectrum in Hu--Sawicki $f(R)$ gravity}
\label{subsec:multipoles}

\begin{figure*}
\centering
\subfloat[]{\includegraphics[width=0.45\textwidth]{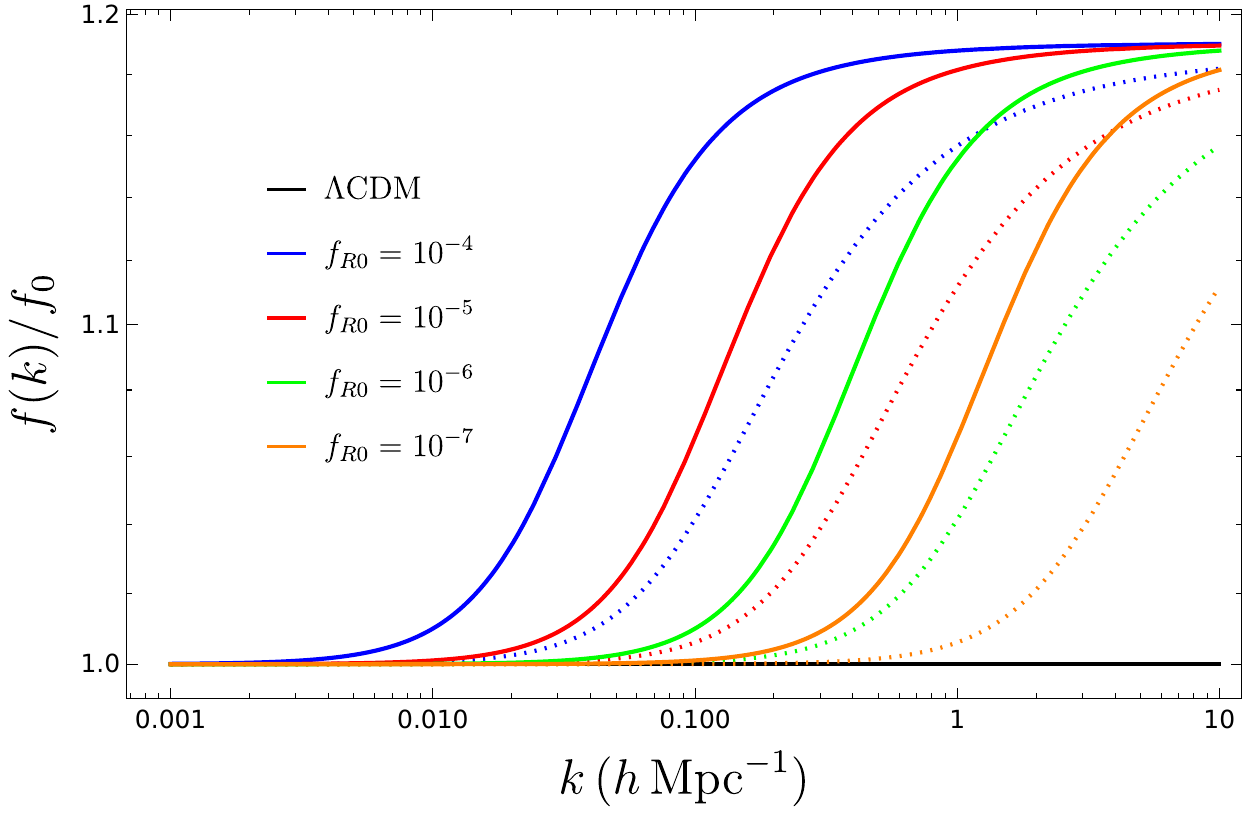}}
\subfloat[]{\includegraphics[width=0.45\textwidth]{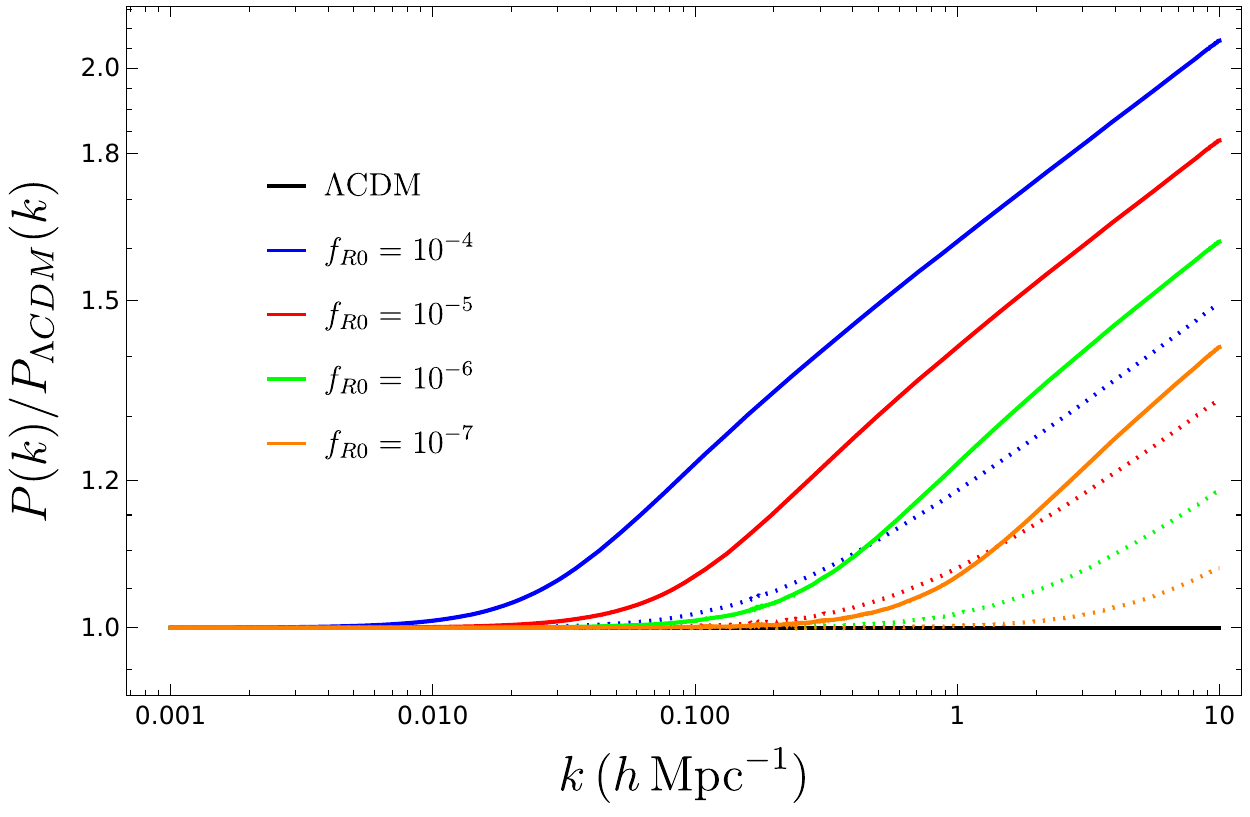}}
\caption{\justifying Scale dependence of linear growth in Hu--Sawicki $f(R)$ gravity. \textit{Left:} the logarithmic growth rate $f(k)$ normalized by its large-scale value $f_0\equiv f(k=10^{-4}\,h\,\mathrm{Mpc}^{-1})$, shown for several $|f_{R0}|$ values. \textit{Right:} the ratio of the linear matter power spectrum in $f(R)$ to that in $\Lambda$CDM for the same parameters. In both panels, solid and dotted curves correspond to $z=0$ and $z=2$, respectively. Deviations from $\Lambda$CDM increase toward smaller scales (larger $k$), while they are reduced at higher redshift. These trends reflect the Yukawa-like enhancement of gravity below the Compton wavelength in HS-$f(R)$ and motivate searching for modified-gravity signatures in quasi-linear and small-scale clustering statistics.}

\label{fig:fk_pk}
\end{figure*}

\begin{figure*}
\centering
\subfloat[]{\includegraphics[width=0.32\textwidth]{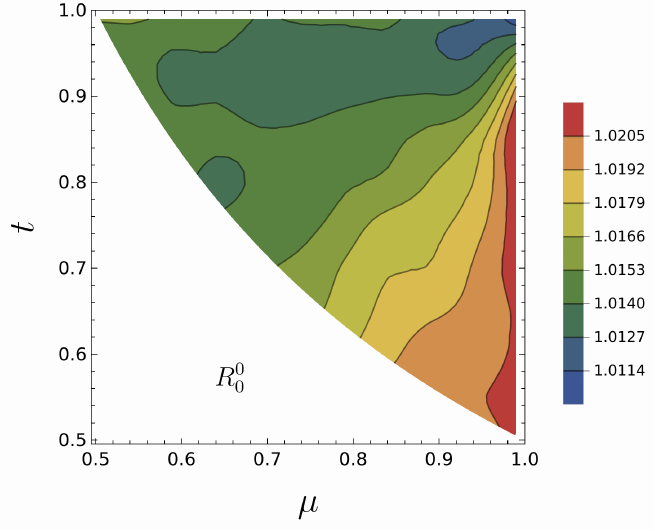}}
\subfloat[]{\includegraphics[width=0.32\textwidth]{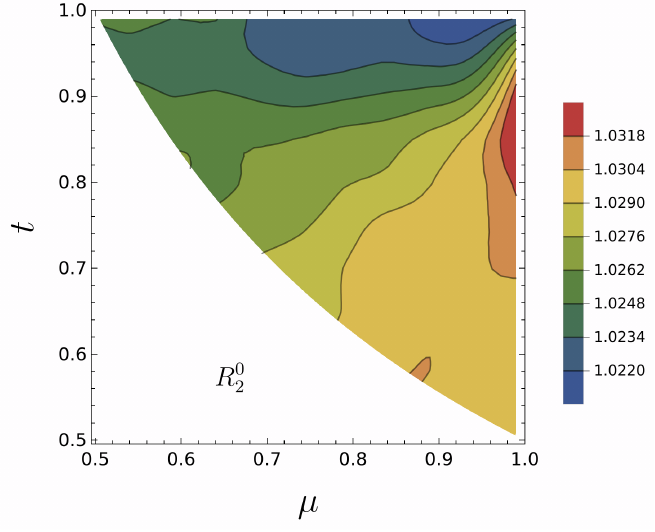}}
\subfloat[]{\includegraphics[width=0.32\textwidth]{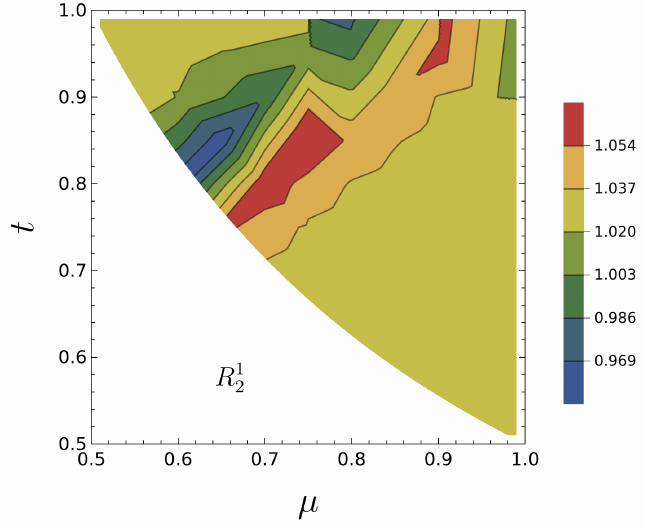}}\\

\subfloat[]{\includegraphics[width=0.32\textwidth]{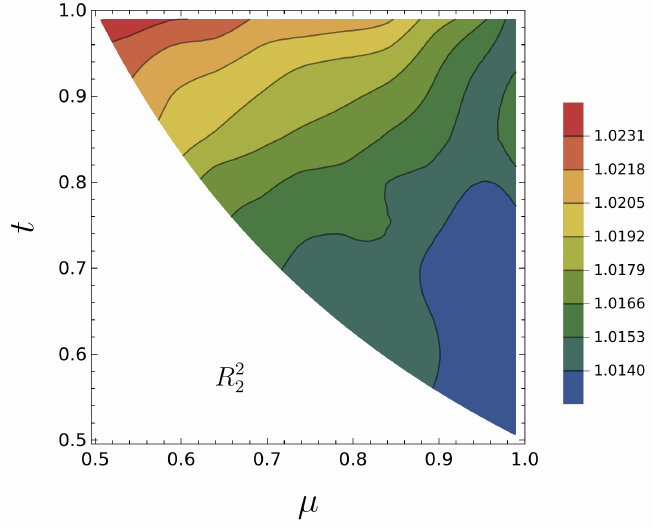}}
\subfloat[]{\includegraphics[width=0.32\textwidth]{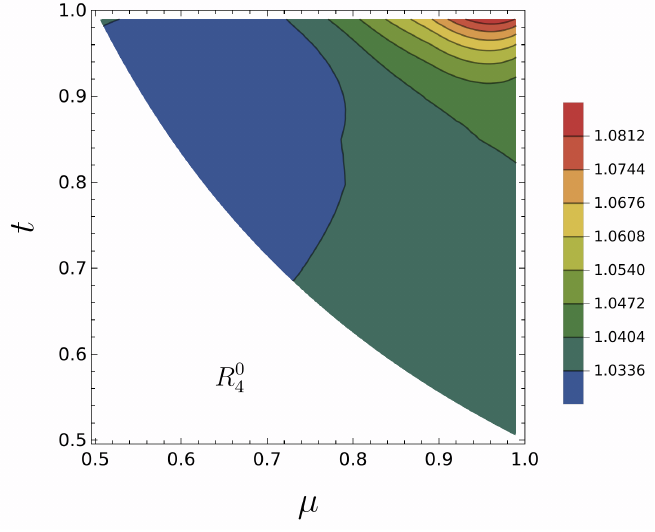}}
\subfloat[]{\includegraphics[width=0.32\textwidth]{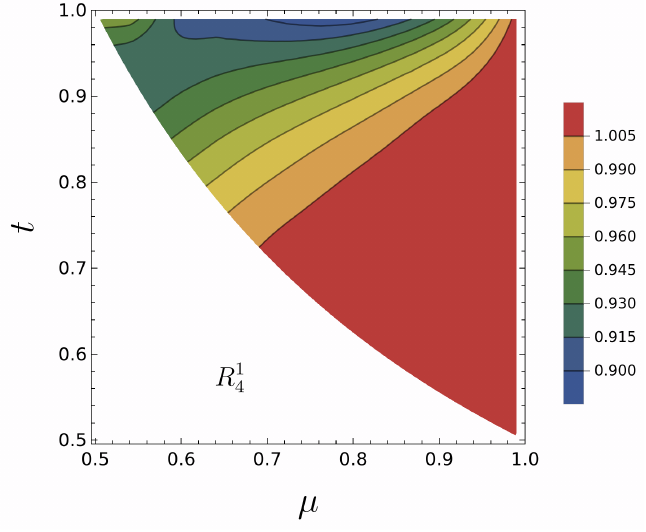}}
\caption{\justifying Ratios of redshift-space \HI\ bispectrum multipoles in Hu--Sawicki $f(R)$ gravity relative to $\Lambda$CDM, evaluated at $z=2$ and $k_1=0.1\,h\,\mathrm{Mpc}^{-1}$ for $|f_{R0}|=10^{-5}$. Each heatmap shows
$R_\ell^m(k_1,\mu,t)\equiv B_\ell^m|_{\rm HS}/B_\ell^m|_{\Lambda{\rm CDM}}$
across the allowed $(\mu,t)$ triangle domain. The lowest multipoles exhibit coherent percent-level departures, with the monopole and quadrupole ratios enhanced most strongly toward nearly linear configurations. Higher-order multipoles display more distinctive configuration dependence: in particular, the $\ell=4$ modes show localized regions of larger fractional deviations, including both enhancement and suppression depending on $(\mu,t)$.}

\label{fig:ratio_multipoles_HS}
\end{figure*}

The matter power spectrum in the Hu--Sawicki model has been studied extensively (e.g.\ Refs.~\cite{Rodriguez-Meza:2023rga,Aviles:2018saf,Aviles:2017aor}), and galaxy-bispectrum analyses have also been developed in this framework~\citep{Aviles:2023fqx,Pal:2025zep}. For 21-cm observables, $f(R)$ effects on the power spectrum have been explored in~\cite{Dash:2024jfi,Dash:2020yfq}, and post-EoR bispectrum forecasts have been presented in \cite{Durrer:2020orn,Jolicoeur:2020eup,Karagiannis:2022ylq}. In this section, we focus on the redshift-space \HI\ bispectrum multipoles in the HS-$f(R)$ model during the post-reionization epoch.

As discussed in Sec.~\ref{seubsec:beyond_GR}, the HS model enhances clustering on sufficiently small scales while recovering the $\Lambda$CDM behavior on large scales. Fig.~\ref{fig:fk_pk} illustrates this scale dependence through the logarithmic growth rate and the linear matter power spectrum.\footnote{The linear power spectrum for the HS-$f(R)$ model (for the corresponding value of $f_{R0}$) is generated using the public code \texttt{fkpt}: \href{https://github.com/alejandroaviles/fkpt}{https://github.com/alejandroaviles/fkpt}.}
In the left panel we show the scale-dependent growth rate $f(k)$, normalized by its large-scale value $f_0=f(k=10^{-4}\,h\,{\rm Mpc}^{-1})$, which matches $\Lambda$CDM. Solid (dotted) lines correspond to $z=0$ ($z=2$), and the black curve denotes $\Lambda$CDM. The right panel shows the ratio of the linear matter power spectrum in $f(R)$ gravity to that in $\Lambda$CDM for the same redshifts and model parameters. At $z=0$, deviations increase with $k$ and reach the level of $\sim 10$--$20\%$ in $f(k)$ for $f_{R0}\in[10^{-7},10^{-4}]$, while at $z=2$ the deviation starts at higher $k$ and the overall enhancement is reduced.

We compute the tree-level redshift-space \HI\ bispectrum in HS-$f(R)$ model using the same expression as in $\Lambda$CDM (Eq.~\ref{eq:B_rsd}), but with the modified second-order density and velocity kernels given by Eqs.~\ref{eq:F2_HS} and \ref{eq:G2_HS}. To quantify departures from $\Lambda$CDM at the level of individual multipoles, we define
\begin{equation}
R_\ell^m(k_1,\mu,t)\;\equiv\;\frac{B_\ell^m(k_1,\mu,t)\big|_{\rm HS}}{B_\ell^m(k_1,\mu,t)\big|_{\Lambda{\rm CDM}}}\,.
\end{equation}
Fig.~\ref{fig:ratio_multipoles_HS} shows $R_\ell^m$ at $k_1=0.1\,h\,{\rm Mpc}^{-1}$ and $z=2$ across the allowed $(\mu,t)$ plane, for $f_{R0}=10^{-5}$.

For the monopole, $R_0^0$ exhibits a percent-level enhancement, ranging from $\sim 1.1\%$ to $2.0\%$, with the largest deviations occurring toward nearly linear configurations. The $(\ell=2,m=0)$ multipole shows a similar overall pattern but with a slightly larger and more anisotropic response, reaching deviations up to $\sim 3.2\%$ in parts of configuration space. In contrast, the $(\ell=2,m=1)$ multipole displays both enhancement and suppression at the level of $\sim 4$--$5\%$, indicating that azimuthal structure can be more sensitive to the interplay between the modified kernels and redshift-space anisotropies.

The lower panels of Fig.~\ref{fig:ratio_multipoles_HS} show that the configuration dependence becomes more distinctive for $m\neq 0$ and for higher $\ell$. The $(\ell=2,m=2)$ ratio reaches deviations of order a few percent, with enhancements at the level of $\sim 2\%$--$3\%$ in the equilateral configuration and a smaller enhancement of $\sim 1.4\%$ in stretched triangles. The most pronounced departure among the multipoles shown appears in the $(\ell=4,m=0)$ mode: regions near the squeezed limit reach nearly an $8\%$ enhancement, and most of the $(\mu,t)$ plane deviates at the level of $\gtrsim 3\%$. The $(\ell=4,m=1)$ mode exhibits a pronounced sign-asymmetric response: for $\mu>t$ the ratio is mildly enhanced but remains small ($\sim 0.5\%$), whereas for $\mu<t$ the multipole is suppressed, with the maximum suppression occurring near the L-isosceles configuration at the level of $\sim 10\%$.

Overall, Fig.~\ref{fig:ratio_multipoles_HS} shows that HS-$f(R)$ effects manifest as coherent percent-level shifts in the lowest multipoles, together with more localized (and sometimes larger) deviations in higher-order multipoles. In the next section, we assess the observational impact of these signatures for an \texttt{SKA-MID}-like survey.

\section{SNR forecasts for \texttt{SKA-MID}}
\label{sec:snr}
In practice, the \HI\ intensity mapping observations are affected by instrumental and thermal noise, which must be incorporated in the covariance of the bispectrum estimator. In what follows, we model the noise properties of \texttt{SKA-MID} by closely following the treatments in Refs.~\citep{cosmicVisions21cm:2018rfq,Dash:2024jfi,Yadav:2024kzh,Dash:2023scq,Weltman:2018zrl}, and we adopt the covariance-matrix methodology developed in the previous analyses \citep{Mazumdar:2020bkm,Pal:2025hpl,Pal:2025zep}. In this formalism, the instrumental contribution enters through the thermal-noise power spectrum, $P_N$, while the sampling of Fourier configurations is captured by the number of independent triangles, $N_{\rm tr}$, within the survey volume.

\subsection{Thermal-noise power spectrum}
\label{subsec:thermal_noise}

The dominant instrumental contribution for a 21-cm interferometric survey arises from the thermal noise of the receiver system. We consider an array of $N_{\rm ant}=250$ identical dishes of diameter $D_{\rm dish}=15\,{\rm m}$ and aperture efficiency $\eta=0.7$, for which the effective collecting area per antenna is
\begin{equation}
A_e=\eta\,\frac{\pi D_{\rm dish}^2}{4}\simeq 123.7\,{\rm m}^2.
\end{equation}
For a Fourier mode with transverse component $k_\perp=k\sqrt{1-(\hat{\mathbf{k}}\cdot \hat{\mathbf{n}})^2}$, the corresponding interferometer baseline length is
\begin{equation}
b=\frac{\lambda\,k_\perp\,r(z)}{2\pi},
\end{equation}
where $\lambda=0.21(1+z)\,{\rm m}$ is the observed wavelength of the redshifted 21-cm line, and $r(z)$ is the comoving distance to redshift $z$.

We approximate the baseline distribution as $\rho(b)\propto 1/b^2$ between a minimum and maximum baseline set by the dish diameter and the array size (here $\approx1\,{\rm km}^2$), respectively. The distribution is normalized such that the total number of visibility pairs is correctly accounted for. Under these assumptions, the effective integration time for a mode is
\begin{equation}
t_k=\frac{T_{\rm obs}\,N_{\rm ant}(N_{\rm ant}-1)\,A_e\,\rho(b)}{2\lambda^2},
\end{equation}
where we take $T_{\rm obs}=400\,{\rm hrs}$ per pointing (with multiple pointings used for wide-area coverage) and a bandwidth $B=32\,{\rm MHz}$ corresponding to observations around $z=2$. We assume that the system temperature is dominated by the sky contribution and adopt $T_{\rm sys}=60\,{\rm mK}$ at $z=2$.

\begin{figure*}
\centering
\subfloat[]{\includegraphics[width=0.32\textwidth]{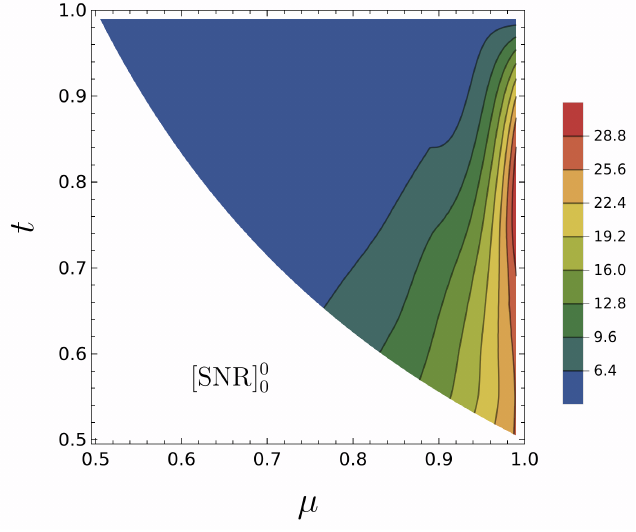}}
\subfloat[]{\includegraphics[width=0.32\textwidth]{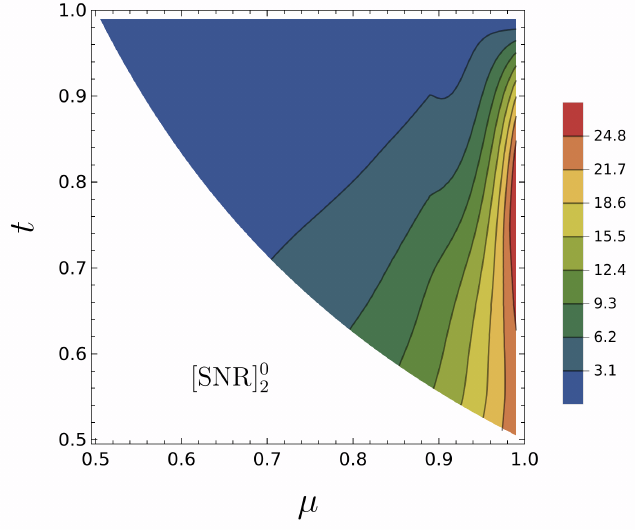}}
\subfloat[]{\includegraphics[width=0.32\textwidth]{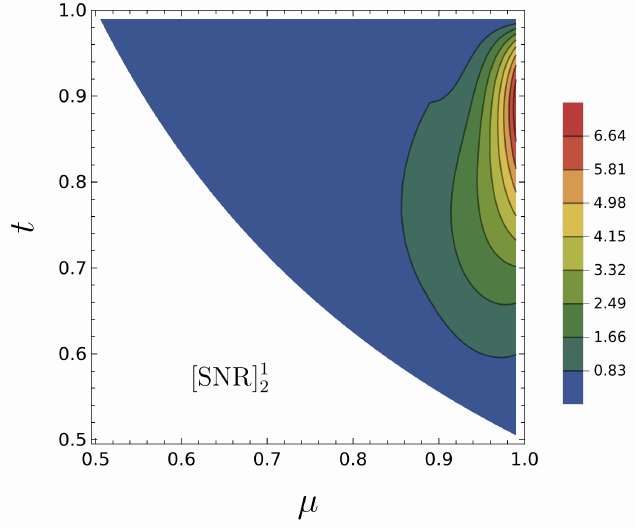}}\\
\subfloat[]{\includegraphics[width=0.32\textwidth]{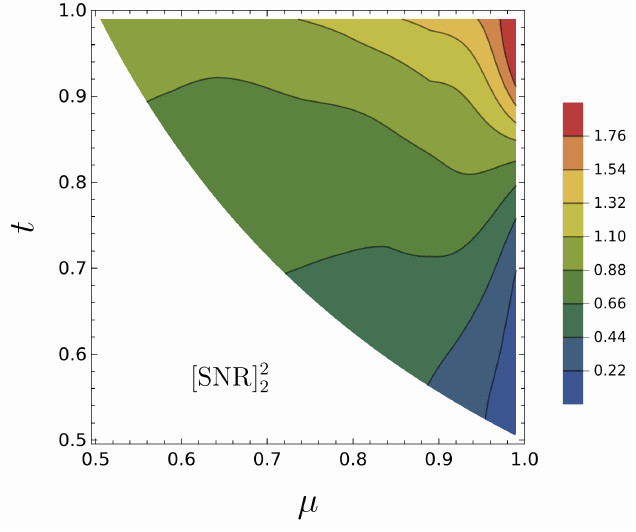}}
\subfloat[]{\includegraphics[width=0.32\textwidth]{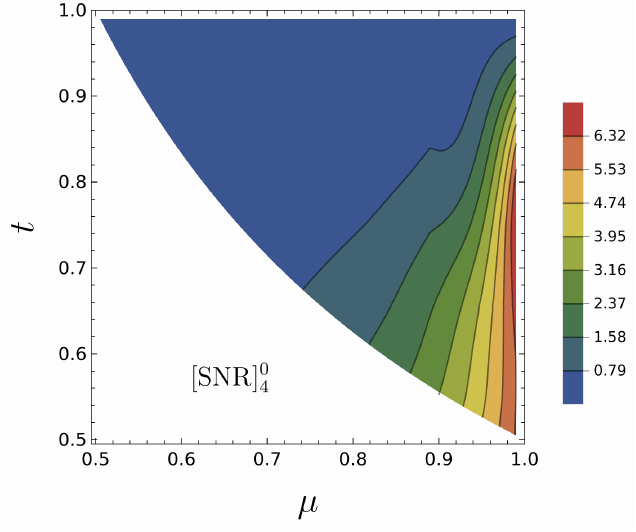}}
\subfloat[]{\includegraphics[width=0.32\textwidth]{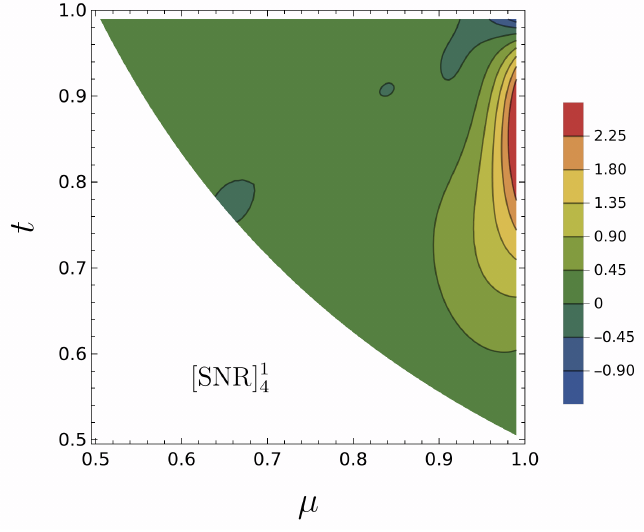}}
\caption{\justifying Signal-to-noise ratio (SNR) maps for modified-gravity signatures in redshift-space \HI\ bispectrum multipoles for an \texttt{SKA-MID}-like intensity-mapping survey, shown in the $(\mu,t)$ plane at $z=2$ and $k_1=0.1\,h\,\mathrm{Mpc}^{-1}$ for HS-$f(R)$ with $|f_{R0}|=10^{-5}$. The signal is defined as the multipole difference
$\Delta B_\ell^m \equiv B_\ell^m|_{\rm HS}-B_\ell^m|_{\Lambda{\rm CDM}}$,
and the noise is given by the diagonal covariance $C_{\ell\ell}^{mm}$ computed from Eq.~\eqref{eq:cov}. The detectability is concentrated toward nearly linear configurations ($\mu\to 1$), with stretched and squeezed triangles providing the highest SNR, while equilateral and L-isosceles configurations are strongly suppressed. The overall SNR is dominated by the $m=0$ multipoles, with azimuthal modes ($m=1,2$) contributing significantly only in localized regions near the high-$\mu$ boundary.}

\label{fig:snr_multipoles}
\end{figure*}

The thermal-noise power spectrum per mode can then be written as
\begin{equation}
P_N(k,\mu)=\frac{\lambda^2\,T_{\rm sys}^2\,r^2(z)}{A_e\,t_k}\,\frac{dr}{d\nu},
\end{equation}
where $dr/d\nu$ converts frequency intervals into comoving radial distance. We also define the $\mu$-averaged noise power spectrum,
\begin{equation}
P_N(k)=\langle P_N(k,\mu)\rangle_\mu,
\end{equation}
which is used when an orientation-averaged noise model is required.

\subsection{Bispectrum multipole estimator and number of triangles}
\label{subsec:bispec_estimator_snr}

Following the multipole-based bispectrum estimation methodology developed in Refs.~\citep{Pal:2025hpl,Mazumdar:2022ynd}, we apply the same formalism to the redshift-space \HI\ bispectrum. The estimator for the bispectrum multipoles is defined as a weighted sum over all triangle triplets $(\mathbf{k}_n,\bar{\mathbf{k}}_n,\tilde{\mathbf{k}}_n)$ that close within the bin $(k_1,\mu,t)$:
\begin{equation}
\label{eq:bispec_estimator}
\begin{split}
\hat{B}_{\ell}^m(k_1,\mu,t)=
\sum_{\text{triplets}}
\frac{w_{\ell}^m(\hat{\mathbf{p}}_n)}{2V}
\Big[
&\delta_{\mathrm{HI}}^s(\mathbf{k}_n)\,
\delta_{\mathrm{HI}}^s(\bar{\mathbf{k}}_n)\,
\delta_{\mathrm{HI}}^s(\tilde{\mathbf{k}}_n)
\\
&+\text{c.c.}
\Big] .
\end{split}
\end{equation}
where ``c.c.'' denotes the complex conjugate, and $V$ is the comoving survey volume associated with a single redshift bin.

Rather than defining the radial depth through the instrumental bandwidth, we adopt fixed redshift bins of width $\Delta z=0.1$ to provide a uniform treatment across redshift and to simplify comparisons between different epochs. For a bin centered at $z$, the survey volume is
\begin{equation}
V(z)=\frac{4\pi}{3}\,f_{\rm sky}\left[\chi^3(z_{\rm max})-\chi^3(z_{\rm min})\right],
\label{eq:Vsurvey_21cm}
\end{equation}
where $\chi(z)$ is the comoving radial distance and $z_{\rm max}=z+\Delta z/2$, $z_{\rm min}=z-\Delta z/2$. We fix the sky fraction to $f_{\rm sky}=0.48$, consistent with a wide-area \texttt{SKA-MID}-like post-EoR intensity-mapping survey.\footnote{Detailed \texttt{SKA} configuration specifications are available at \href{https://www.skao.int/en/science-users/118/ska-telescope-specifications}{https://www.skao.int/en/science-users/118/ska-telescope-specifications}.}

The total number of independent triangles contributing to a bin centered at $(k_1,\mu,t)$ is denoted $N_{\rm tr}$. With the normalization convention for the harmonic weights used in Refs.~\citep{Mazumdar:2022ynd,Pal:2025hpl,Pal:2025zep}, one has
\begin{equation}
\sum_{n_1}\left|Y_\ell^m(\hat{\mathbf{p}}_{n_1})\right|^2=\frac{4\pi}{N_{\rm tr}},
\end{equation}
with
\begin{equation}
N_{\rm tr}=
\left(\frac{V\,k_1^3}{8\pi^4}\right)^2
t^2\,
\Big[
\Delta\ln k_1\,\big(t\,\Delta\ln k_1+\Delta t\big)\,\Delta\mu
\Big].
\end{equation}
The dependence on $V$ therefore directly controls the sampling variance by setting the number of available triangle configurations.

\subsection{Covariance of bispectrum multipoles}
\label{subsec:covariance_snr}

\begin{figure*}
\centering
\subfloat[]{\includegraphics[width=0.35\textwidth]{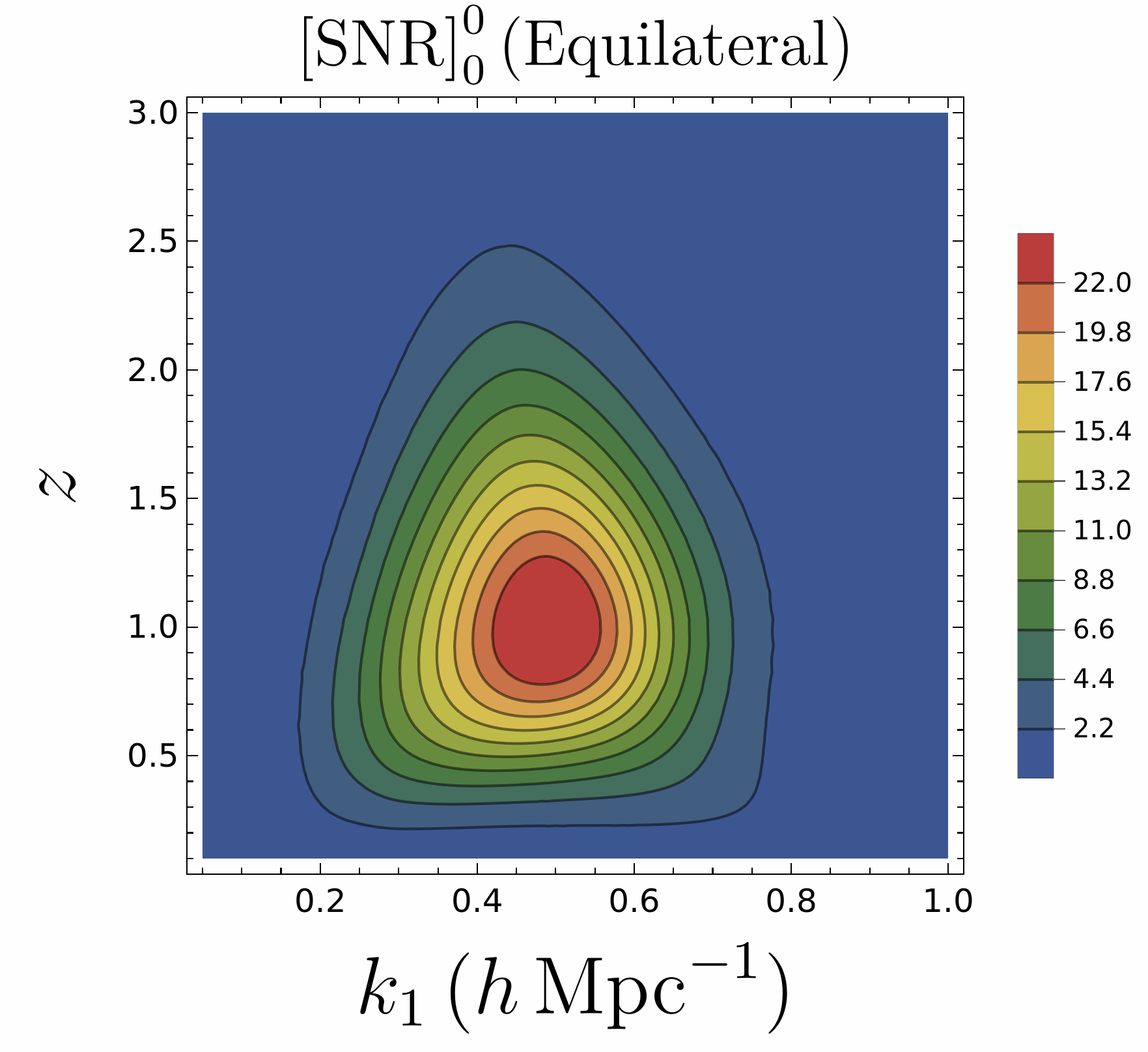}}
\subfloat[]{\includegraphics[width=0.35\textwidth]{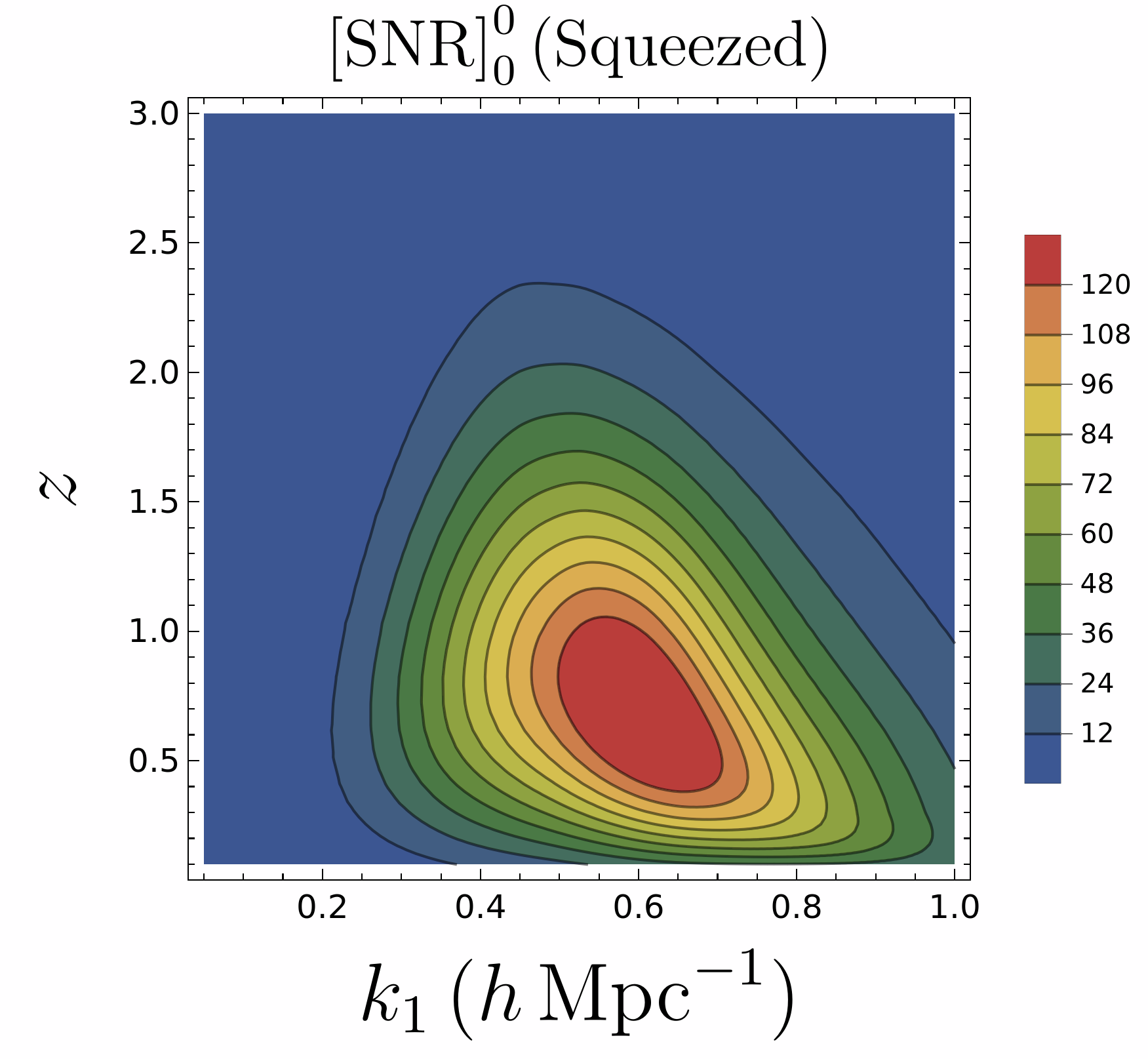}}
\subfloat[]{\includegraphics[width=0.35\textwidth]{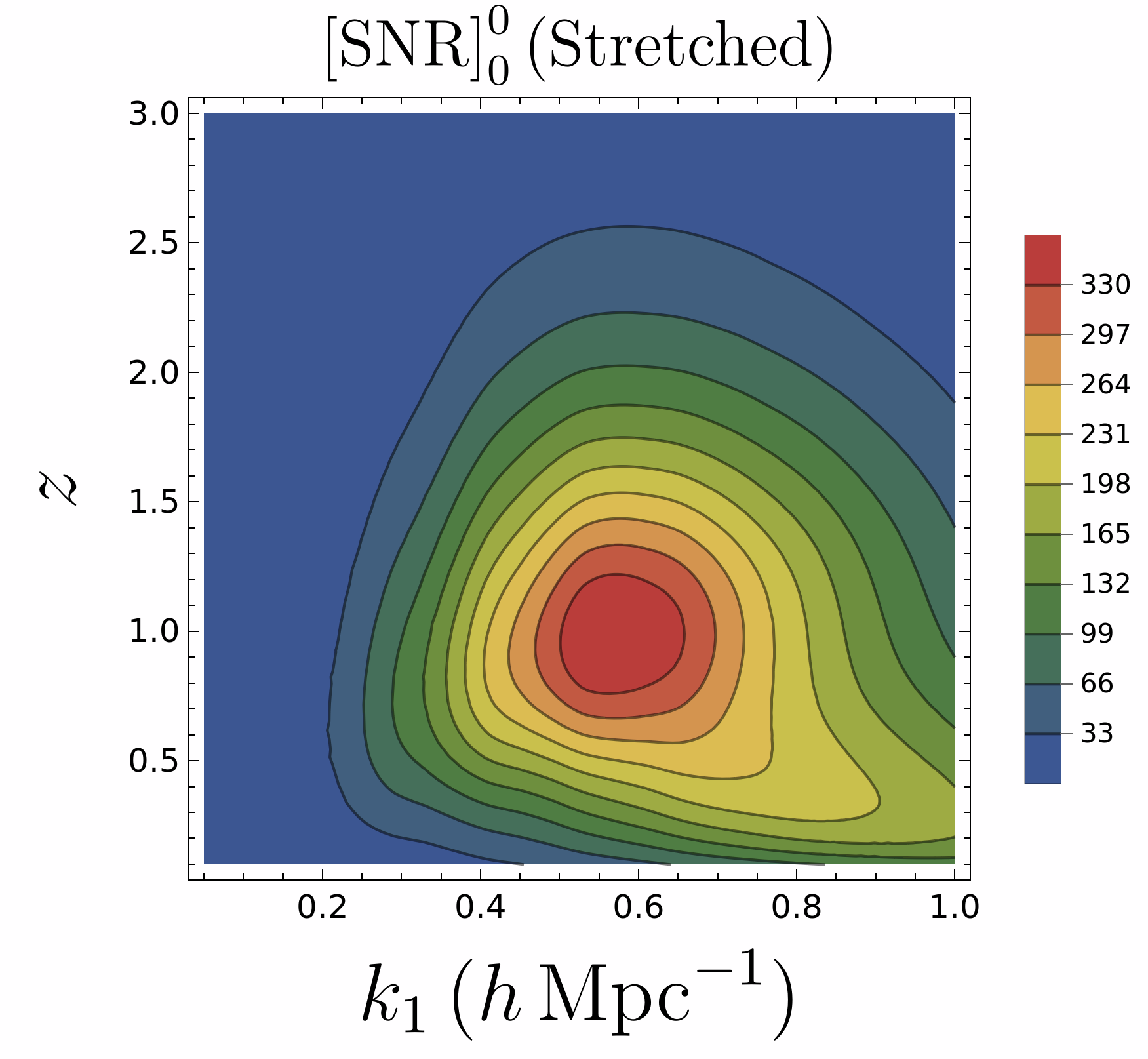}}\\
\subfloat[]{\includegraphics[width=0.35\textwidth]{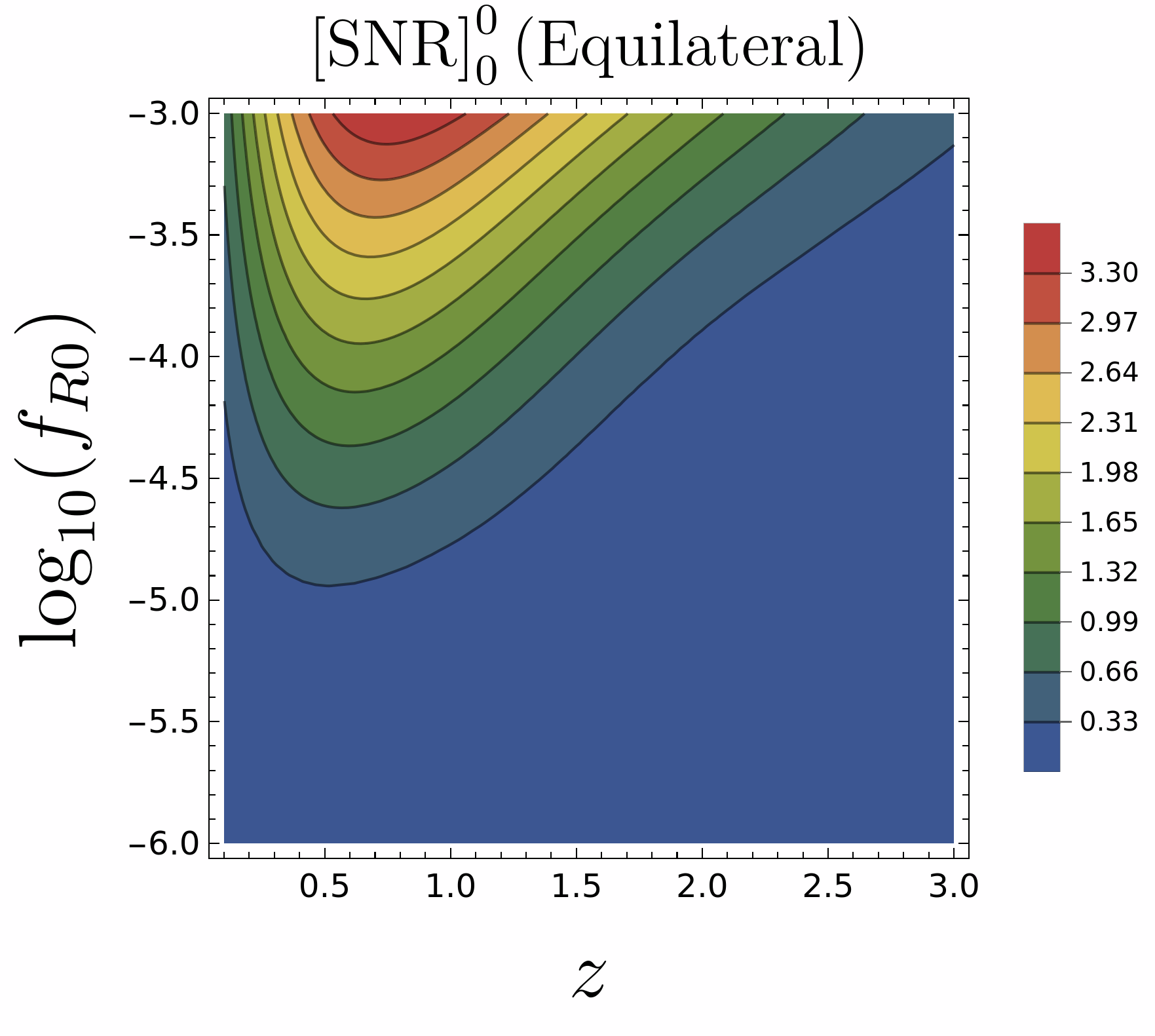}}
\subfloat[]{\includegraphics[width=0.35\textwidth]{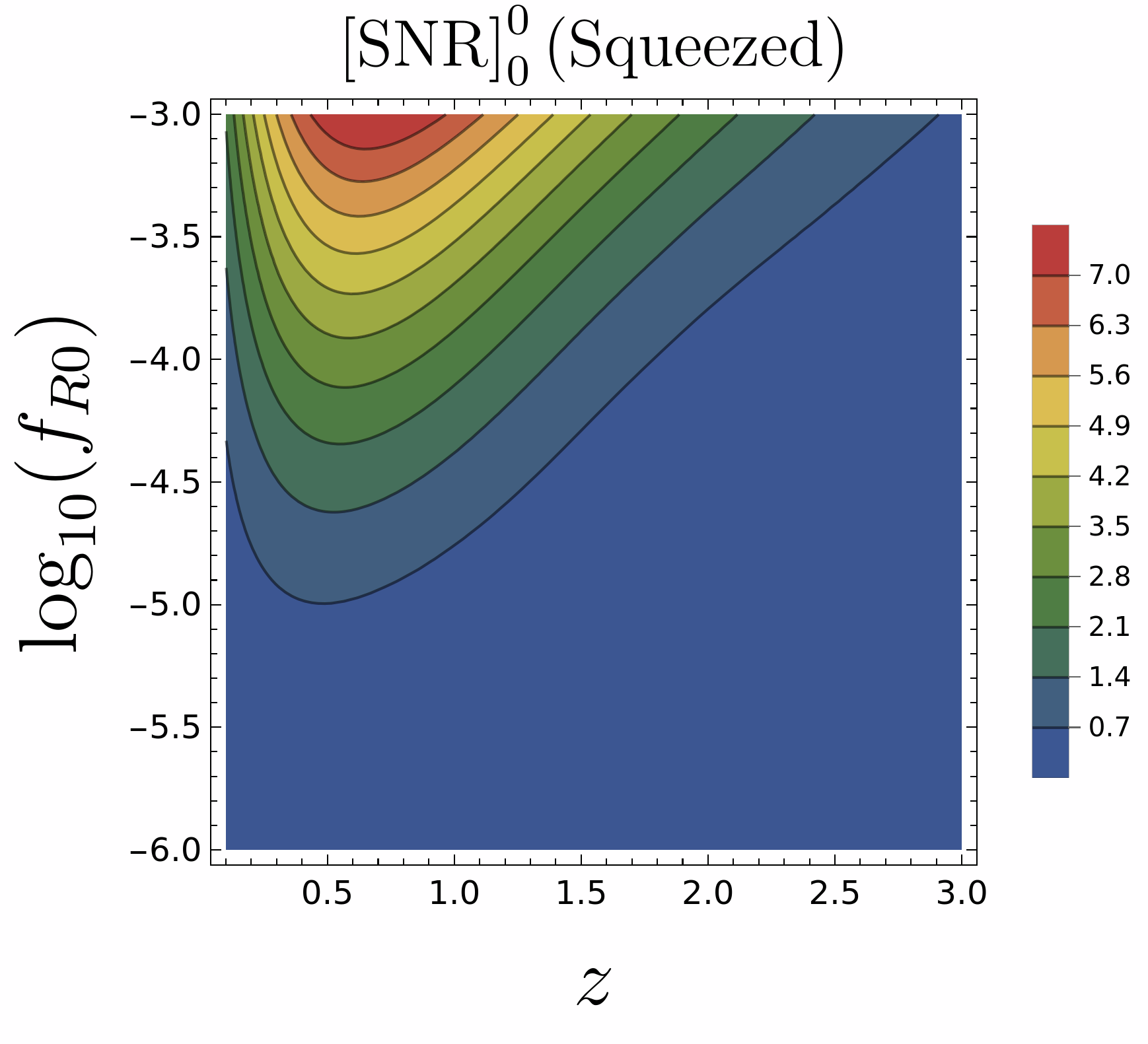}}
\subfloat[]{\includegraphics[width=0.35\textwidth]{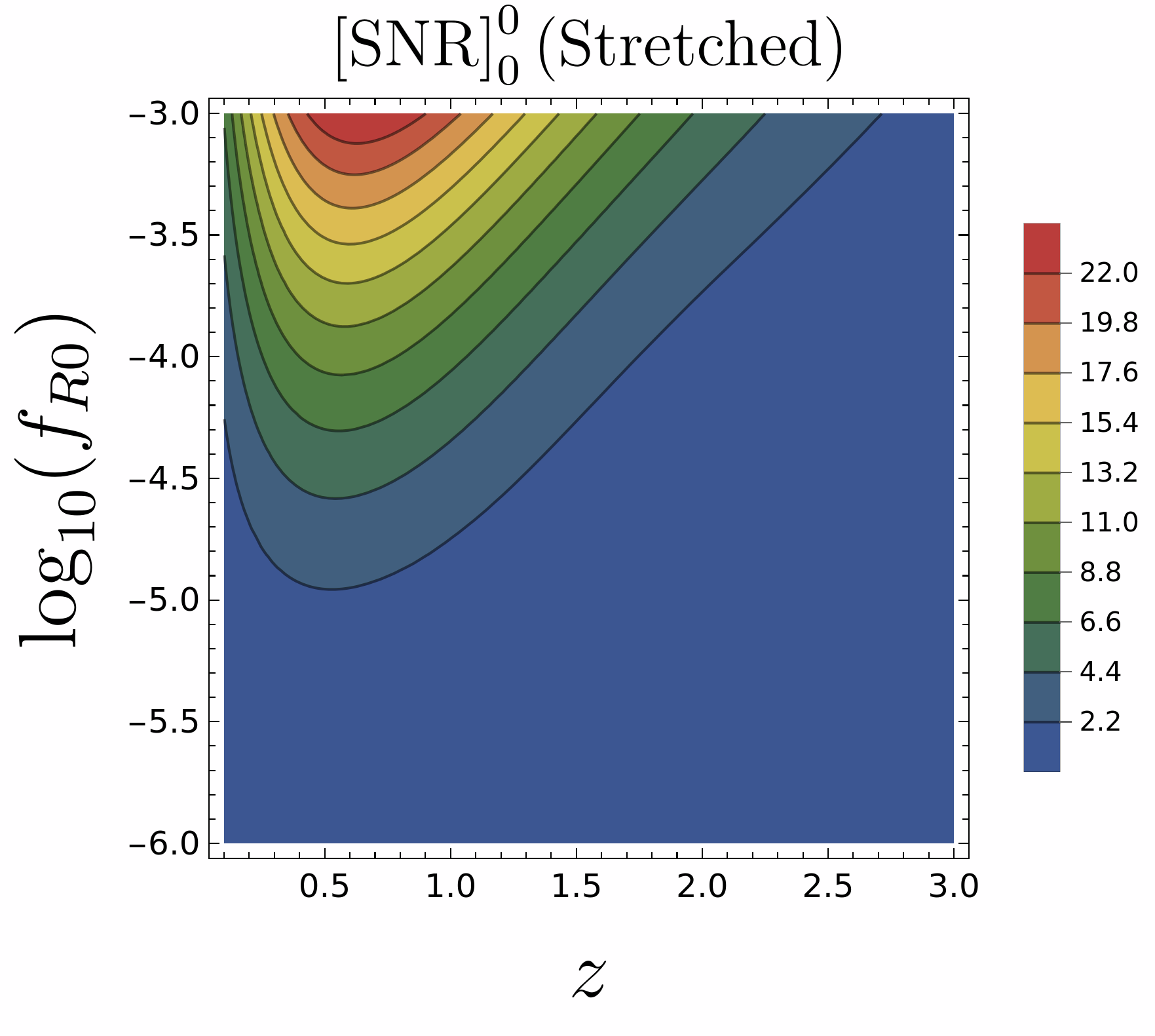}}
\caption{\justifying Redshift and scale dependence of the SNR of the modified-gravity induced \HI\ bispectrum monopole in HS-$f(R)$ gravity for an \texttt{SKA-MID}-like survey, shown for three representative triangle shapes: equilateral (left), squeezed (middle), and stretched (right). \textit{Top row:} SNR in the $(k_1,z)$ plane at fixed $|f_{R0}|=10^{-5}$, illustrating a maximum at intermediate quasi-linear scales where modified-gravity effects are sizable while cosmic variance (low $k_1$) and thermal noise/FoG suppression (high $k_1$) are subdominant. \textit{Bottom row:} SNR in the $(z,\log_{10}|f_{R0}|)$ plane at fixed $k_1=0.1\,h\,\mathrm{Mpc}^{-1}$, showing the monotonic increase of detectability with $|f_{R0}|$ and the decline at higher redshift driven by the rising thermal noise at the corresponding observing frequencies.}

\label{fig:snr_k1_z_multipoles}
\end{figure*}

The covariance matrix of the bispectrum multipole estimator includes contributions from cosmic variance and from thermal noise. The covariance element $C_{\ell\ell'}^{mm'}(k_1,\mu,t)$ can be written as
\begin{align}
C_{\ell \ell'}^{m m'}(k_1,\mu,t)
&=
\frac{\sqrt{(2\ell+1)(2\ell'+1)}}{N_{\rm tr}}
\int d\Omega_{\hat{\mathbf{p}}}
\mathrm{Re}[Y_{\ell}^m]
\mathrm{Re}[Y_{\ell'}^{m'}]
\nonumber\\
&\hspace{-6.6em}\times
\Big[
\underbrace{3\left(B^s(k_1,\mu,t,\hat{\mathbf{p}})\right)^2
+V\,\prod_{i=1}^{3}P_{\mathrm{HI}}^s(k_i,\mu_i)}_{\text{cosmic variance}}
+
\underbrace{V\,\prod_{i=1}^{3}P_N(k_i,\mu_i)}_{\text{thermal noise}}
\Big].
\label{eq:cov}
\end{align}
The first term inside the brackets arises from the cosmic variance of the bispectrum, while the second term includes contributions from the instrumental thermal noise through $P_N$.

\section{Results}
\label{subsec:snr_results}

In this section, we present signal-to-noise ratio (SNR) forecasts for the redshift-space \HI\ bispectrum multipoles, including the full bispectrum covariance and \texttt{SKA-MID} thermal noise. As discussed earlier, the anisotropic bispectrum admits non-vanishing multipoles up to $\ell\le 8$ and $m\le 6$, but higher-order multipoles are strongly suppressed and contribute negligibly to the total detectability. We therefore focus on the dominant modes up to $(\ell,m)=(4,1)$.

Beyond establishing the configuration-space hierarchy of detectable multipoles, it is essential to identify \emph{where} in scale and redshift the MG-induced signal is maximized relative to the total covariance. This motivates a two-step presentation: we first map the SNR across triangle configurations at a representative redshift and scale, and then study how the same signal evolves with $(k_1,z)$ and with the MG parameter $|f_{R0}|$.

Our baseline predictions are obtained in the fiducial $\Lambda$CDM cosmology, and we quantify MS signatures by considering the difference between the HS-$f(R)$ model and $\Lambda$CDM. For a given triangle bin $(k_1,\mu,t)$, we define the MG-induced bispectrum difference
\begin{equation}
\Delta B_\ell^m(k_1,\mu,t)\equiv 
B_\ell^m(k_1,\mu,t)\big|_{\rm HS}
-
B_\ell^m(k_1,\mu,t)\big|_{\Lambda{\rm CDM}}\,,
\end{equation}
and evaluate the corresponding (configuration-dependent) SNR as
\begin{equation}
\left[{\rm SNR}\right]_\ell^m(k_1,\mu,t)=
\frac{\Delta B_\ell^m(k_1,\mu,t)}
{\sqrt{C_{\ell\ell}^{mm}(k_1,\mu,t)}}\,,
\label{eq:snr_def}
\end{equation}
where $C_{\ell\ell}^{mm}$ is computed from Eq.~\ref{eq:cov}. Throughout this section, the bispectrum multipoles $B_\ell^m$ are obtained from Eq.~\ref{eq:blm} by projecting the redshift-space bispectrum $B^s(k_1,\mu,t)$ (Eq.~\ref{eq:B_rsd}) onto spherical harmonics, and the noise power spectra are modeled for \texttt{SKA-MID} as described in Sec.~\ref{sec:snr}. Unless stated otherwise, we adopt $|f_{R0}|=10^{-5}$ and fix all other cosmological parameters to the fiducial values used in the previous sections.
\begin{figure*}
\centering
\subfloat[]{\includegraphics[width=0.35\textwidth]{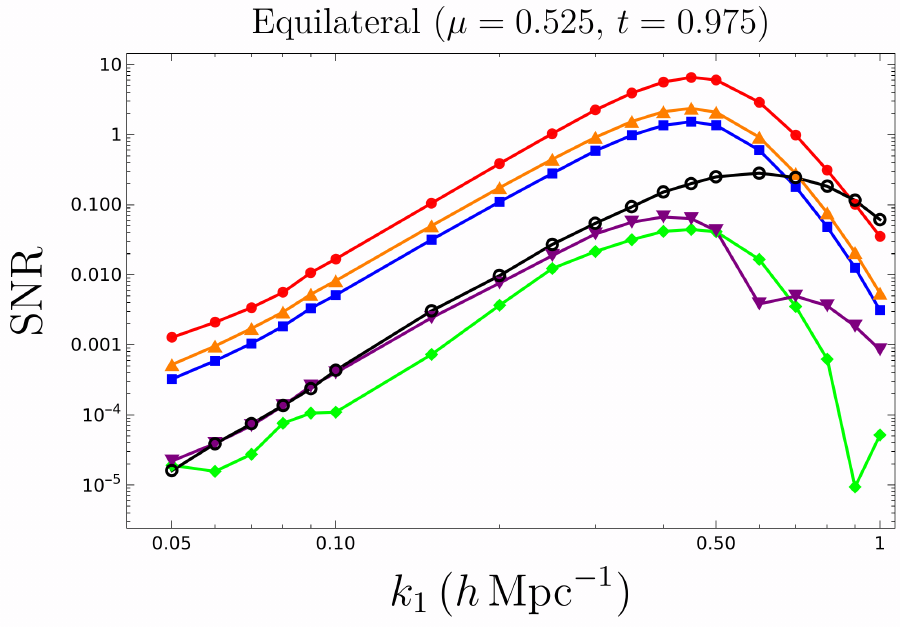}}
\subfloat[]{\includegraphics[width=0.35\textwidth]{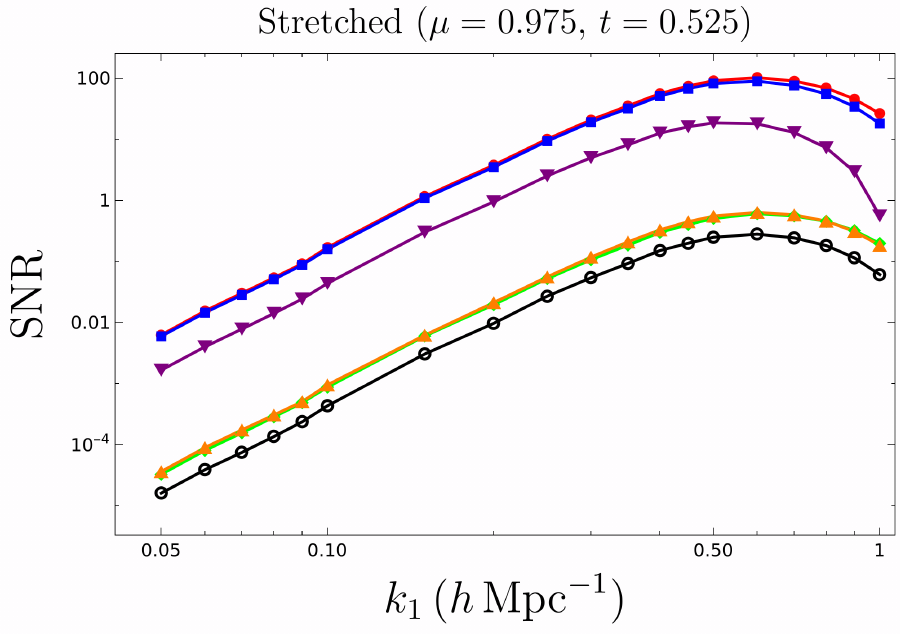}}
\subfloat[]{\includegraphics[width=0.35\textwidth]{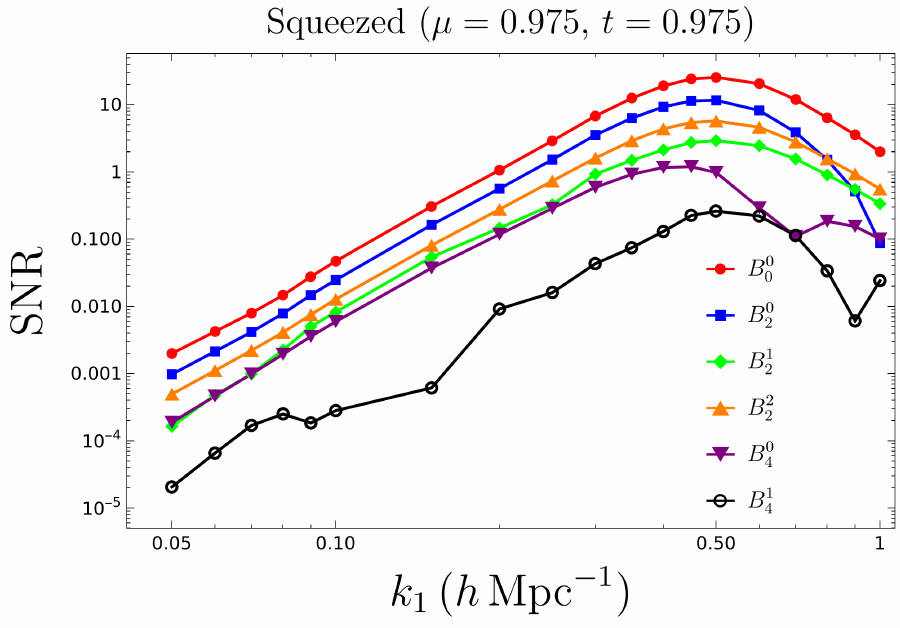}}
\caption{\justifying Scale dependence of the SNR of individual redshift-space \HI\ bispectrum multipoles as a function of the largest triangle side $k_1$, evaluated at fixed redshift $z=2$ for HS-$f(R)$ with $|f_{R0}|=10^{-5}$. We show multipoles up to $\ell=4$ (with the $m$ values indicated in the legend) for three representative triangle shapes: equilateral (left), stretched (middle), and squeezed (right). In all cases, the detectable signal is dominated by the lowest multipoles, with higher-order modes increasingly noise-dominated. The stretched and squeezed configurations yield systematically higher SNR than equilateral triangles across scales, reflecting their stronger sensitivity to redshift-space anisotropies and to the scale-dependent growth characteristic of $f(R)$ gravity, while the SNR turns over at high $k_1$ once thermal noise and FoG suppression become significant.}

\label{fig:snr_k1_multipoles}
\end{figure*}

As a first diagnostic, we fix $z=2$ and $k_1=0.1\,h\,{\rm Mpc}^{-1}$ and map the SNR over the full triangle configuration space. 
Fig.~\ref{fig:snr_multipoles} shows SNR maps in the $(\mu,t)$ plane for multipoles with $\ell=0,2,4$ and $m=0,1,2$, evaluated at $k_1=0.1\,h\,{\rm Mpc}^{-1}$ and $z=2$. Across all panels, the SNR is strongly concentrated near nearly linear triangle configurations ($\mu\to 1$), where redshift-space anisotropies maximize the contrast between MG and GR predictions. By contrast, equilateral and L-isosceles configurations yield significantly lower SNR, consistent with stronger FoG damping and isotropic velocity-dispersion effects in these regions of configuration space.

The monopole $\left[{\rm SNR}\right]_0^0$ provides the dominant contribution, reaching peak values of $\sim 25$--$30$ in the band $t\simeq 0.55$--$0.95$ at $\mu\to 1$. Configurations with $\mu>t$ are particularly sensitive, reflecting the enhanced response of stretched triangles to the scale-dependent growth characteristic of HS-$f(R)$ gravity. $\left[{\rm SNR}\right]_2^0$ follows a similar trend but with slightly reduced amplitude, reaching $\sim 20$--$25$ in the same region. $\left[{\rm SNR}\right]_4^0$ is only marginally detectable, with peak values of $\sim 5$--$6$, again confined to $\mu\to1$.

The azimuthal modes are generally subdominant. The multipoles $\left[{\rm SNR}\right]_2^1$ and $\left[{\rm SNR}\right]_2^2$ reach typical maxima of order $\sim 4$--$7$ and $\sim 1$--$2$, respectively, and are detectable only in localized regions near the high-$\mu$, high-$t$ boundary. Higher-order azimuthal multipoles such as $\left[{\rm SNR}\right]_4^1$ remain below unity across almost the entire $(\mu,t)$ plane except a small patch near the linear configuration, indicating that their MG imprint is subdominant compared to the combined effect of cosmic variance and instrumental noise.

While these $(\mu,t)$ maps identify the most informative triangle shapes at fixed $(k_1,z)$, the overall detectability also depends on \emph{where} the survey noise and cosmic variance are minimized relative to the MG signal, which varies with both scale and redshift. We therefore next explore the SNR dependence on $(k_1,z)$ and on the MG amplitude $|f_{R0}|$ for representative triangle shapes.

To identify the optimal observational window for MG detection, we now examine how the SNR varies with scale and redshift.
Fig.~\ref{fig:snr_k1_z_multipoles} summarizes the redshift and scale dependence of the monopole SNR for three representative triangle shapes (equilateral, squeezed, and stretched). Since the other multipoles exhibit similar qualitative trends (with different amplitudes), we focus on the monopole for clarity. The top row shows the SNR in the $(k_1,z)$ plane at fixed $|f_{R0}|=10^{-5}$. For equilateral configurations, the SNR peaks at intermediate scales, $k_1\simeq 0.4$--$0.6\,h\,{\rm Mpc}^{-1}$, and around $z\simeq 1$, where the MG-induced signal is large compared to the total variance. Squeezed triangles exhibit substantially higher SNR, reaching values up to $\sim 120$, reflecting the enhanced mode coupling in the squeezed limit and its sensitivity to scale-dependent growth. Stretched triangles provide the largest SNR, with peak values approaching $\sim 300$, with the maximum shifting toward slightly higher wavenumbers ($k_1\simeq 0.6$--$1.0\,h\,{\rm Mpc}^{-1}$) and lower redshifts. For all shapes, the SNR decreases at low $k_1$ due to cosmic variance and at high $k_1$ due to the increasing impact of thermal noise and velocity-dispersion effects.

The bottom row of Fig.~\ref{fig:snr_k1_z_multipoles} shows the SNR in the $(z,\log_{10}|f_{R0}|)$ plane at fixed $k_1=0.1\,h\,{\rm Mpc}^{-1}$. In all configurations, the SNR increases monotonically with $|f_{R0}|$. Equilateral triangles provide the weakest sensitivity, with a peak SNR of $\sim 3$--$3.5$ near $z\simeq 0.8$. Squeezed configurations reach $\sim 7$ near $z\simeq 0.5$, while stretched configurations yield the strongest constraints, with SNR values of order $\sim 20$ for the same scale. At $z\gtrsim 2$, the SNR drops rapidly for all shapes, driven by the increase in \texttt{SKA-MID} thermal noise at the corresponding observing frequencies.

Having identified the preferred $(k_1,z)$ regions and their dependence on $|f_{R0}|$, we now connect these trends back to the \emph{multipole content} by examining how the SNR is distributed across individual $(\ell,m)$ modes as a function of scale for representative triangle shapes.

Fig.~\ref{fig:snr_k1_multipoles} complements this picture by illustrating, at fixed $z=2$, the scale dependence of the SNR for individual multipoles up to $\ell=4$ (with $m\le 1$) for equilateral, stretched, and squeezed configurations. In the equilateral case, the SNR is dominated by the lowest-order multipoles, with the monopole $B_0^0$ and the quadrupole $B_2^2$ providing the largest contributions and peaking at $k_1\simeq 0.4$--$0.6\,h\,{\rm Mpc}^{-1}$, while higher-order multipoles are strongly suppressed (with $B_2^1$ being the most suppressed among those shown). In the stretched case, the SNR is enhanced across multipoles by nearly an order of magnitude; the quadrupole becomes comparable to the monopole, and even $B_4^0$ attains appreciable SNR at larger $k_1$, whereas $B_2^1$, $B_2^2$, and $B_4^1$ remain subdominant. For squeezed triangles, the SNR of all multipoles increases with $k_1$ up to $k_1\simeq 0.7$--$0.8\,h\,{\rm Mpc}^{-1}$, beyond which thermal noise dominates. Apart from the monopole, multipoles up to $(\ell,m)=(2,2)$ remain statistically significant at large $k_1$, while $\ell=4$ modes provide little additional constraining power.

Taken together, these results show that the most detectable HS-$f(R)$ signatures in the post-reionization 21-cm bispectrum are carried by the lowest multipoles (primarily $\ell=0,2$ with $m=0$), and by triangle configurations close to the squeezed or stretched limits. The highest sensitivity is achieved at intermediate scales, $k_1\simeq 0.4$--$0.8\,h\,{\rm Mpc}^{-1}$, and at redshifts $z\simeq 1$--$2$, where the MG-induced signal is large while thermal noise remains subdominant. Higher-order multipoles are increasingly noise dominated, and equilateral configurations contribute comparatively little, reflecting the combined effects of \texttt{SKA-MID} thermal noise, redshift-space anisotropies, and the scale-dependent growth characteristic of $f(R)$ gravity.

\section{Conclusions}
\label{sec:conclusion}
We have presented a detailed analysis of the redshift-space \HI\ bispectrum multipoles in the post-reionization era, with the dual goal of (i) characterizing the anisotropic, non-linear \HI\ clustering signal in redshift space, and (ii) assessing its potential as a probe of departures from General Relativity. Within a tree-level perturbative framework, we modeled the \HI\ bispectrum in $\Lambda$CDM including local quadratic bias, Kaiser redshift-space distortions, and a phenomenological Fingers-of-God (FoG) damping. We then extended the same formalism to Hu--Sawicki $f(R)$ gravity by consistently incorporating the scale-dependent linear growth and the modified second-order perturbation kernels. To connect theory with observations, we quantified detectability using a full bispectrum covariance model that includes both cosmic variance and thermal noise appropriate for a \texttt{SKA-MID}-like survey.

Our $\Lambda$CDM results show that redshift-space distortions substantially enhance the bispectrum relative to real space and imprint a rich angular dependence that is efficiently captured by a spherical-harmonic multipole decomposition. The signal is dominated by the lowest-order multipoles, particularly the monopole and quadrupole, which retain large amplitudes across a broad region of triangle configuration space. These modes are most prominent for nearly linear, stretched, and squeezed configurations, where coherent line-of-sight velocities maximize the anisotropic mode-coupling. Higher-order multipoles exhibit increasingly intricate angular patterns, but their amplitudes are strongly suppressed and their covariance increases rapidly, rendering them subdominant for practical detection with \texttt{SKA-MID}-level sensitivity.

In Hu--Sawicki $f(R)$ gravity, we find coherent, configuration-dependent departures from $\Lambda$CDM in the bispectrum multipoles at the percent level. The most robust signatures appear in the monopole and quadrupole components, with the largest leverage coming from squeezed and stretched configurations that preferentially couple long- and short-wavelength modes. Although some higher-order multipoles can show larger \emph{fractional} deviations, their larger variance prevents these deviations from translating into comparable observational significance once the full covariance is taken into account.

A key outcome of our signal-to-noise forecasts is that the modified-gravity sensitivity peaks on intermediate to small (quasi-linear) scales, \(k_{1}\sim 0.4\)–\(0.8\,h\,\mathrm{Mpc}^{-1}\), and at redshifts \(z\sim 1\)–\(2\). This behavior is physically expected from the competition between cosmic variance at low \(k\), thermal noise and FoG suppression at high \(k\), and the fact that modified-gravity effects are imprinted most strongly through scale-dependent growth and non-linear mode coupling. Importantly, this scale preference also strengthens the observational relevance of bispectrum-based tests: the largest-scale (small-\(k\)) modes are notoriously difficult to recover in 21-cm analyses because foreground mitigation and instrumental systematics preferentially contaminate or remove low-\(k_{\parallel}\) and ultra-large-scale modes. In contrast, recent progress with current-generation instruments (e.g.\ CHIME and MeerKAT-based intensity-mapping analyses) indicates that quasi-linear scales are increasingly accessible, making the regime where our bispectrum multipoles have the highest forecast SNR particularly timely for near-future surveys.

Taken together, our results highlight the strong potential of \texttt{SKA-MID}-based post-EoR \HI\ intensity-mapping observations to exploit bispectrum multipoles as a sensitive and complementary probe of gravity with 21-cm fluctuations. By compressing the redshift-space anisotropy into a well-defined set of multipoles and leveraging the configuration dependence of non-linear mode coupling, the bispectrum captures information about structure formation that is inaccessible to two-point statistics alone.

These results provide a strong motivation for pushing bispectrum-multipole forecasts toward more realistic modeling and analysis conditions. First, we have worked at tree level with a simplified \HI\ bias model and a phenomenological FoG prescription; improving this will require loop/EFT corrections, more flexible and physically motivated \HI\ biasing, and validation against simulations. Second, our detectability forecasts rely on an idealized noise model and do not include foreground residuals, mode-mixing from realistic survey window functions, calibration/beam systematics, or anisotropic filtering effects that can couple to the multipole estimator. Incorporating these effects---together with tomographic multi-redshift analyses and multi-tracer cross-correlations---will be essential for converting the promising signal-to-noise levels found here into robust constraints on $f(R)$ gravity and other theories beyond $\Lambda$CDM.

\section*{Acknowledgement}
\label{sec:conclusion}
SP acknowledges CSIR for financial support through Senior Research Fellowship (File no. 09/093(0195)/2020-EMR-I). 
DS acknowledges the support of the Canada 150 Chairs program, the Fonds de recherche du Québec Nature et Technologies (FRQNT) and the Natural Sciences and Engineering Research Council of Canada (NSERC) joint NOVA grant, and the Trottier Space Institute Postdoctoral Fellowship program.

\begin{figure*}
\centering
  \subfloat[]{\includegraphics[width=0.35\textwidth]{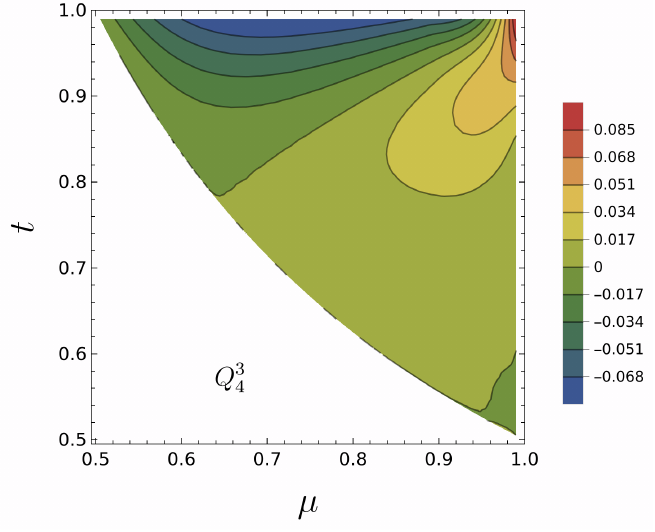}}
  \subfloat[]{\includegraphics[width=0.35\textwidth]{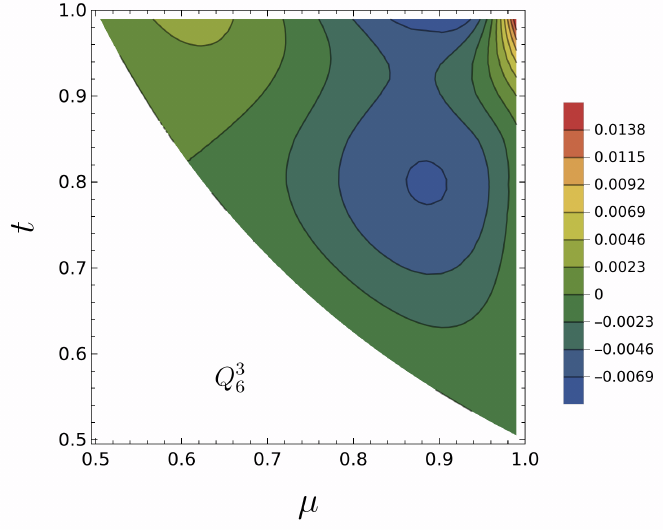}}
  \subfloat[]{\includegraphics[width=0.35\textwidth]{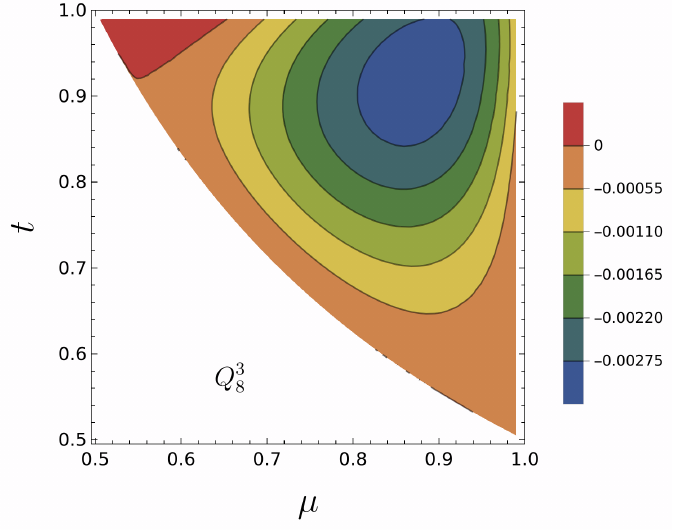}}\\
  \subfloat[]{\includegraphics[width=0.35\textwidth]{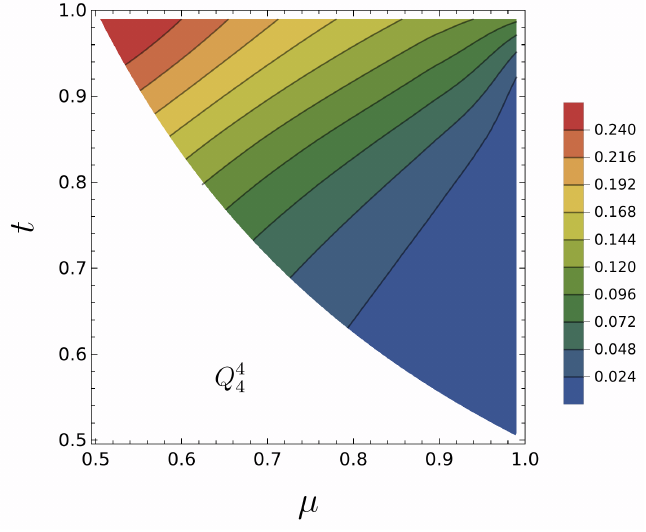}}
  \subfloat[]{\includegraphics[width=0.35\textwidth]{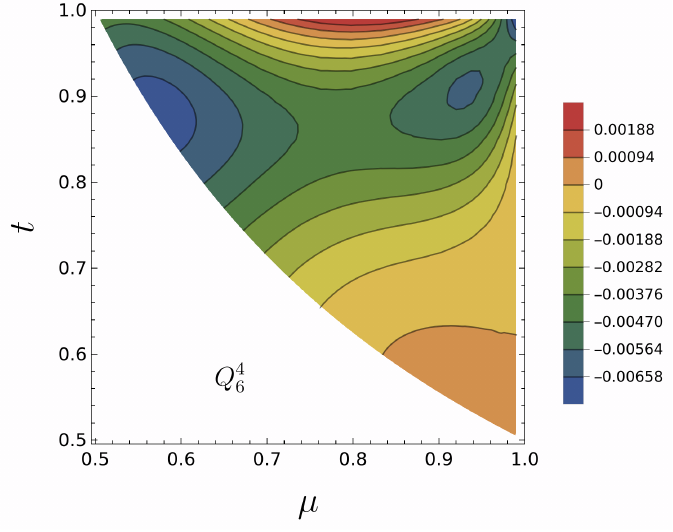}}
  \subfloat[]{\includegraphics[width=0.35\textwidth]{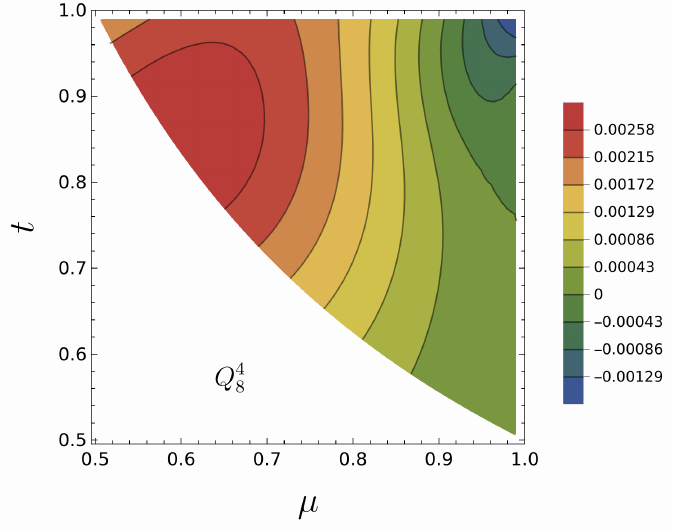}}
\caption{\justifying Reduced redshift-space \HI\ bispectrum multipoles $Q_\ell^m(\mu,t)$ in $\Lambda$CDM at $k_1=0.1\,h\,\mathrm{Mpc}^{-1}$ and $z=2$, shown for higher azimuthal modes $m=3$ (top row) and $m=4$ (bottom row) and for $\ell=4,6,8$ (left to right). These higher-$m$ multipoles are strongly suppressed compared to the lowest-order modes and exhibit alternating positive and negative regions across the allowed triangle domain, reflecting increasingly oscillatory angular projections. The dominant features tend to localize toward the squeezed/near-linear boundary, while the overall amplitudes decrease rapidly with increasing $\ell$, consistent with strong angular cancellations.}

\label{fig:Qm3m4_multipoles}
\end{figure*}

\bibliographystyle{apsrev4-2}
 \bibliography{references}  

\appendix
\section{Higher-order multipoles in $\Lambda$CDM}
\label{app:higher_multipoles}

In this appendix, we present the behaviour of higher-order redshift-space bispectrum multipoles
in a fiducial $\Lambda$CDM cosmology for completeness, although they are not expected to be observationally dominant.
These multipoles are expected to be strongly suppressed compared to the real-space bispectrum,
but nevertheless exhibit non-trivial angular dependence that encodes anisotropic redshift-space effects.

\begin{figure*}[t]
\centering
\subfloat[]{\includegraphics[width=0.35\textwidth]{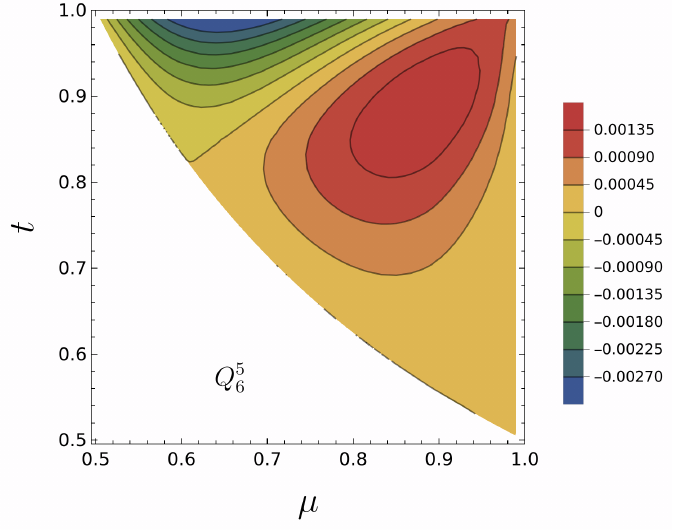}}
\subfloat[]{\includegraphics[width=0.35\textwidth]{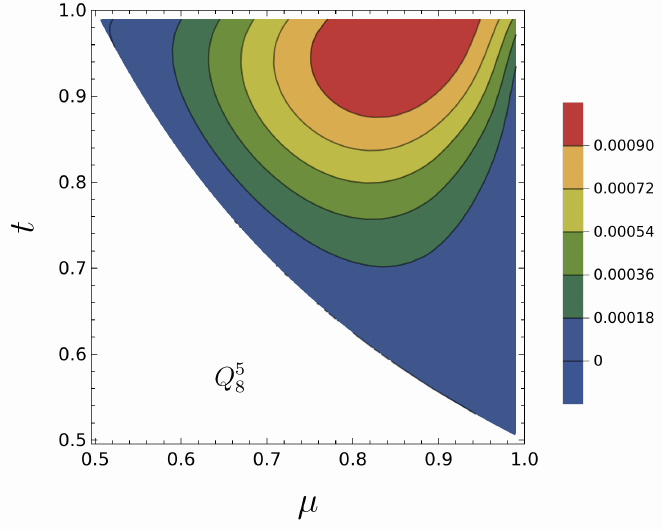}}\\
\subfloat[]{\includegraphics[width=0.35\textwidth]{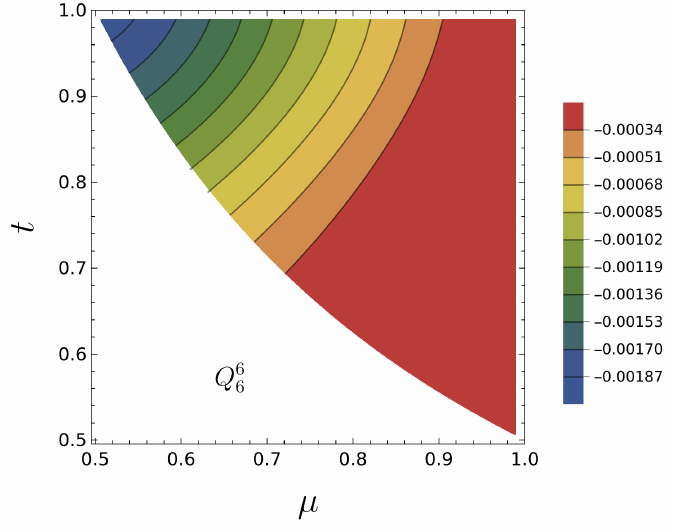}}
\subfloat[]{\includegraphics[width=0.35\textwidth]{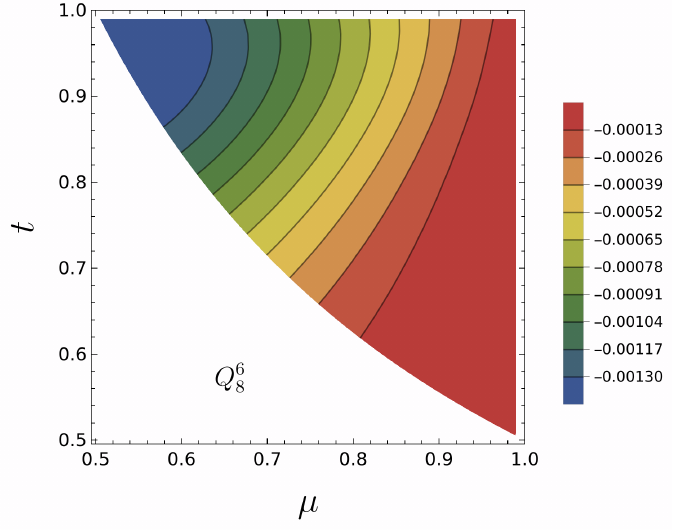}}

\caption{\justifying Same as Fig.~\ref{fig:Qm3m4_multipoles}, but for multipoles with $m=5$ (upper panel) and $m=6$ (lower panel).
All these multipoles are suppressed by approximately $3$--$5$ orders of magnitude relative to the real-space
bispectrum.
The $B_6^5$ multipole is positive over most of the configuration space except near the equilateral limit,
whereas $B_8^5$ is nearly vanishing except in a small region close to the squeezed configuration.
For $m=6$, the bispectrum is predominantly negative, with the largest contributions arising in the $\mu > t$
region of the configuration space.
}
\label{fig:Qm5m6_multipoles}
\end{figure*}

\end{document}